\documentclass[1p,onecolumn,authoryear,preprint,times,10pt]{elsarticle-ed}

\journal{arXiv}

\usepackage[english]{babel}
\usepackage[T1]{fontenc}
\usepackage[utf8]{inputenc}
\usepackage{amsmath,amssymb}
\usepackage{fullpage}
\usepackage{booktabs}
\usepackage[dvipsnames]{xcolor}
\usepackage{soul}

\usepackage{type1cm}   			 
\usepackage{lineno}    			 
\usepackage{lipsum}
\usepackage{subfig}  			 
\usepackage{bm}        			 

\newcommand{\dsum}[0]{\displaystyle\sum}               

\newcommand{\field}[1]{\mathbb{#1}}                    

\defcitealias{2019YoungQ}{Part I}

\begin{document}

\begin{frontmatter}

\title{\textbf{\LARGE Shadowing the rotating annulus. Part II:\\Gradient descent in the perfect model scenario}}

\author[lse,aopp]{\textbf{Roland~M.~B.~Young}\corref{cor1}}
\author[lse]{\textbf{Roman~Binter}}
\author[lse,bios]{\textbf{Falk~Nieh\"{o}rster}}
\author[aopp]{\textbf{Peter L. Read}}
\author[lse,pembroke]{\textbf{Leonard A. Smith}}

\address[lse]{Centre for the Analysis of Time Series, London School of Economics, London, UK}
\address[aopp]{Atmospheric, Oceanic and Planetary Physics, Department of Physics, University of Oxford, Oxford, UK}
\address[bios]{Bermuda Institute of Ocean Sciences, St. George's, Bermuda}
\address[pembroke]{Pembroke College, Oxford, UK}

\cortext[cor1]{Corresponding author. Current address: College of Science, UAE University, P.O. Box 15551, Al Ain, United Arab Emirates. \textit{Email address:} roland.young@uaeu.ac.ae.}

\begin{abstract}
Shadowing trajectories are model trajectories consistent with a sequence of observations of a system, given a distribution of observational noise. The existence of such trajectories is a desirable property of any forecast model. Gradient descent of indeterminism is a well-established technique for finding shadowing trajectories in low-dimensional analytical systems. Here we apply it to the thermally-driven rotating annulus, a laboratory experiment intermediate in model complexity and physical idealisation between analytical systems and global, comprehensive atmospheric models. We work in the perfect model scenario using the MORALS model to generate a sequence of noisy observations in a chaotic flow regime. We demonstrate that the gradient descent technique recovers a pseudo-orbit of model states significantly closer to a model trajectory than the initial sequence. Gradient-free descent is used, where the adjoint model is set to $\lambda{\bf I}$ in the absence of a full adjoint model. The indeterminism of the pseudo-orbit falls by two orders of magnitude during the descent, but we find that the distance between the pseudo-orbit and the initial, true, model trajectory reaches a minimum and then diverges from truth. We attribute this to the use of the $\lambda$-adjoint, which is well suited to noise reduction but not to finely-tuned convergence towards a model trajectory. We find that $\lambda=0.25$ gives optimal results, and that candidate model trajectories begun from this pseudo-orbit shadow the observations for up to 80\,s, about the length of the longest timescale of the system, and similar to expected shadowing times based on the distance between the pseudo-orbit and the truth. There is great potential for using this method with real laboratory data.

$\,$\\
\noindent \textbf{This paper was originally prepared for submission in 2011; but, after Part I was not accepted, it was not submitted. It has not been peer-reviewed. We no longer have the time or resources to work on this topic, but would like this record of our work to be available for others to read, cite, and follow up.\\}

\end{abstract}


\end{frontmatter}

\section{Introduction}

Shadowing trajectories are model trajectories consistent with a sequence of observations of a system, given the distribution of observational noise. The existence of such trajectories is a desirable property of any forecast model. If a model does not admit such trajectories then there is no initial condition that remains close to the observations.

The time over which weather and climate models can shadow observations of past weather and climate is unknown. This is concerning given the weight in decision-making that is placed upon output from these models. Techniques for finding shadowing trajectories are, to some extent, understood in low-dimensional systems such as the \citet{1963Lorenz} equations and the \citet{1979Ikeda} map \citep{2001Judd,2003Judd,2009Du,2010Smith}. There is significant interest in their application to high-dimensional situations such as General Circulation Models (GCMs).

Gradient descent of indeterminism \citep{2003Judd,2008JuddA,2009Stemler} is one such technique well-established for finding shadowing trajectories in low-dimensional analytical systems. It starts from a sequence of observations and alters this sequence by ``descending'' towards a model trajectory. Alterations to each state in the sequence are calculated based on mismatches (forecast errors) between that state and model forecasts forwards and backwards in time from adjacent states mapped on to the state of interest. 

Each state in a sequence constructed by gradient descent is known as a \textit{shadow analysis} \citep{2008JuddA}, and in practice this sequence will be a pseudo-orbit of the model \citep{1975Bowen} rather than a trajectory. These shadow analyses serve as initial conditions for \textit{candidate} model trajectories. There is no \textit{a priori} guarantee that these candidate trajectories will shadow the original observations, but work with analytical systems has shown gradient descent to be a good method for finding such shadowing trajectories in very low dimensional systems \citep[for example]{2010Smith}. The underlying reason for this is not yet well understood. The maximum time that, among all possible candidates, the model shadows the observations for is called the \textit{shadowing time} for that model-observation pair. In practice there are several definitions of shadowing time relevant in different contexts. The $\iota$-shadowing time \citep[p.47]{1998Gilmour} is the maximum time the distance from model trajectory to observations remains within a bound given by observational error, whereas a $\phi$-shadowing time requires the match between model and observations only to be ``useful'' \citep[p.52]{2000SmithA}. In its original sense shadowing refers to the time a true solution of a differential equation remains within a fixed distance of a numerical solution \citep{1975Bowen}. While superficially similar, the distinction is fundamental and we do not consider this so-called $\epsilon$-shadowing in this work.

Gradient descent itself is a method of noise reduction that has been used in nonlinear and chaotic systems for many years \citep{1988Kostelich,1990Grebogi,1990Hammel,1991Farmer}. It has only recently been applied to higher-dimensional systems, by \citet{2004JuddB} using an idealised quasi-geostrophic model and by \citet{2008JuddA} using the US Navy NOGAPS weather model. In neither case were the results used to measure shadowing times against observational data. Its ability to find candidate trajectories that shadow observations for a long time in high-dimensional models of real systems is not yet well explored. Gradient descent has several theoretical advantages over methods in current operational use such as 4D-Var and various flavours of the Kalman filter \citep{2009Stemler,2010Judd}. In low-dimensional test cases it has performed favourably against the extended Kalman filter \citep{2003Judd}, 4D-Var \citep{2009Stemler}, and particle filters \citep{2009Judd}. Its main disadvantage is computational; it is an iterative procedure and $O(100)$ passes through the sequence by the model are required during gradient descent.

\begin{figure}
  \begin{center}
      \includegraphics[viewport = 175 80 420 330, clip,width=0.45\textwidth]{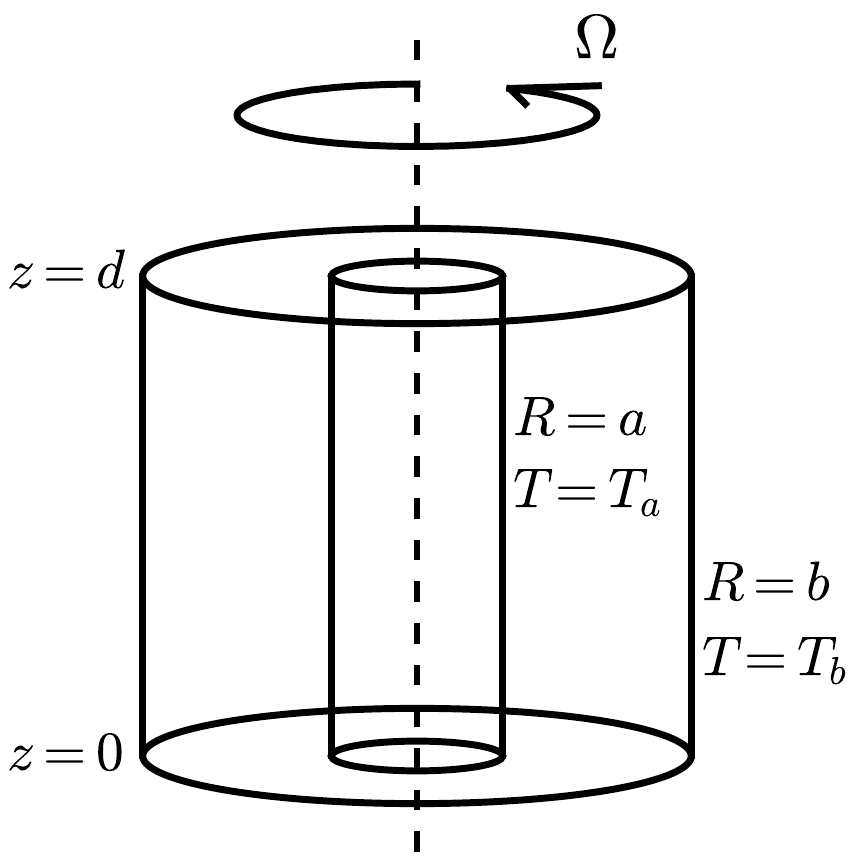}
    \caption{Schematic of a rotating annulus experiment used in the AOPP fluid dynamics laboratory. The inner and outer cylinders at radii $R=a,\,b$ are at temperatures $T_a$ and $T_b$ respectively. The apparatus rotates at constant angular velocity $\Omega$, and has fluid between the two cylinders.}
    \label{fig:annulus-schem}
  \end{center}
\end{figure}

Laboratory experiments are intermediate in model complexity and physical idealisation between analytical and global systems. The laboratory setting allows investigation of the properties of gradient descent using a real physical system and a non-idealised model in a situation where the complexity of the flow can be controlled, the experiments can be repeated, and there is potential for long-range observations under laboratory conditions. Whereas an analytical system may have $O(10)$ variables and a general circulation model of the Earth's atmosphere may have $O(10^7)$, models of laboratory experiments have a more manageable $O(10^4-10^5)$ variables. The thermally-driven rotating annulus (Fig.~\ref{fig:annulus-schem}) is a classic laboratory experiment representing the mid-latitudes of an idealised generic planetary atmosphere. The ``standard'' setup uses two cylinders mounted on a turntable, with coincident axes of rotation. Fluid fills the space between the cylinders, which are enclosed by two water baths. Hot water (relative to the working fluid) is circulated around the outside of the outer cylinder as a heat source, and cold water is circulated around the inside of the inner cylinder as a heat sink. The turntable is rotated, usually anticlockwise. This setup mimics the three major influences acting on a planet's atmosphere: the effects of rotation, gravity, and the temperature difference between low and high latitudes. The annulus exhibits a wide range of dynamical flow regimes describing quasi-periodic, chaotic and turbulent flow, and has become well-established over 50 years as a good laboratory analogue for certain kinds of atmospheric phenomena \citep{1953Hide,1975Hide,1992ReadA}.

Since its early development in the 1950s the annulus has been used to conduct research into the fundamental physical processes underlying weather and climate. In recent years effort has also been directed towards using it to inform the development of methods used for weather and climate forecasting. Under laboratory conditions, properties of a particular method can be studied in isolation but using a real fluid as opposed to idealised analytical models more commonly used when testing new methods. There are also several advantages of the laboratory setting compared with atmospheric studies: the controlled nature of the experiment, the degree of reproducibility of the results, and the avoidance of many of the problems associated with atmospheric observations such as a geographically variable observational data density. Effort so far has been directed towards the application of data assimilation techniques such as analysis correction \citep{2013Young}, the ensemble Kalman filter \citep{2010Ravela}, and the breeding method for ensemble prediction \citep{2015Young}. With a tangent linear and adjoint model of an annulus model one would also be able to test more recent methods for data assimilation such as 4D-Var \citep{2007Rawlins}.

Whether a technique such as gradient descent is feasible for use with high-dimensional GCMs can be informed by its study under the controlled laboratory conditions provied by the annulus experiment. Gradient descent is not yet well-established as a practical method for state estimation in atmospheric systems, but by examining its performance in a real but idealised system a better understanding of whether it could be used operationally will be obtained. A timely comparison to make would be with the results obtained by \citet{2013Young} using the well established analysis correction method.

In this paper we demonstrate the gradient descent technique using a model of the rotating annulus under the controlled conditions afforded by the perfect model scenario. We explore how the results depend on the major tuneable parameter in the algorithm, and calculate shadowing times from candidate trajectories produced by gradient descent using the definition presented in \citet[hereafter \citetalias{2019YoungQ}]{2019YoungQ}. In the future we intend to extend the work to laboratory data, and compare how long our model shadows reality using gradient descent compared with other assimilation methods. \citet{1998Gilmour} attempted to shadow temperature measurements of the annulus using a radial basis function model (but not using gradient descent), but this would be the first attempt to do so using a ``full'' model of this experiment.

The paper is arranged as follows. In Sect.~\ref{sec:annulus} we describe our simulation of the rotating annulus. In Sect.~\ref{sec:gd} the gradient descent method is described and its application to the rotating annulus situation is detailed. Section~\ref{sec:results} shows the results from our perfect model experiments, and shadowing times are calculated in Sect.~\ref{sec:shadowing}. The results are discussed and conclusions are drawn in Sect.~\ref{sec:conc}.

\section{The rotating annulus model}
\label{sec:annulus}

\begin{figure}[tb]
  \begin{center}
	\includegraphics[height=0.45\textwidth,clip,viewport=30 80 420 520]{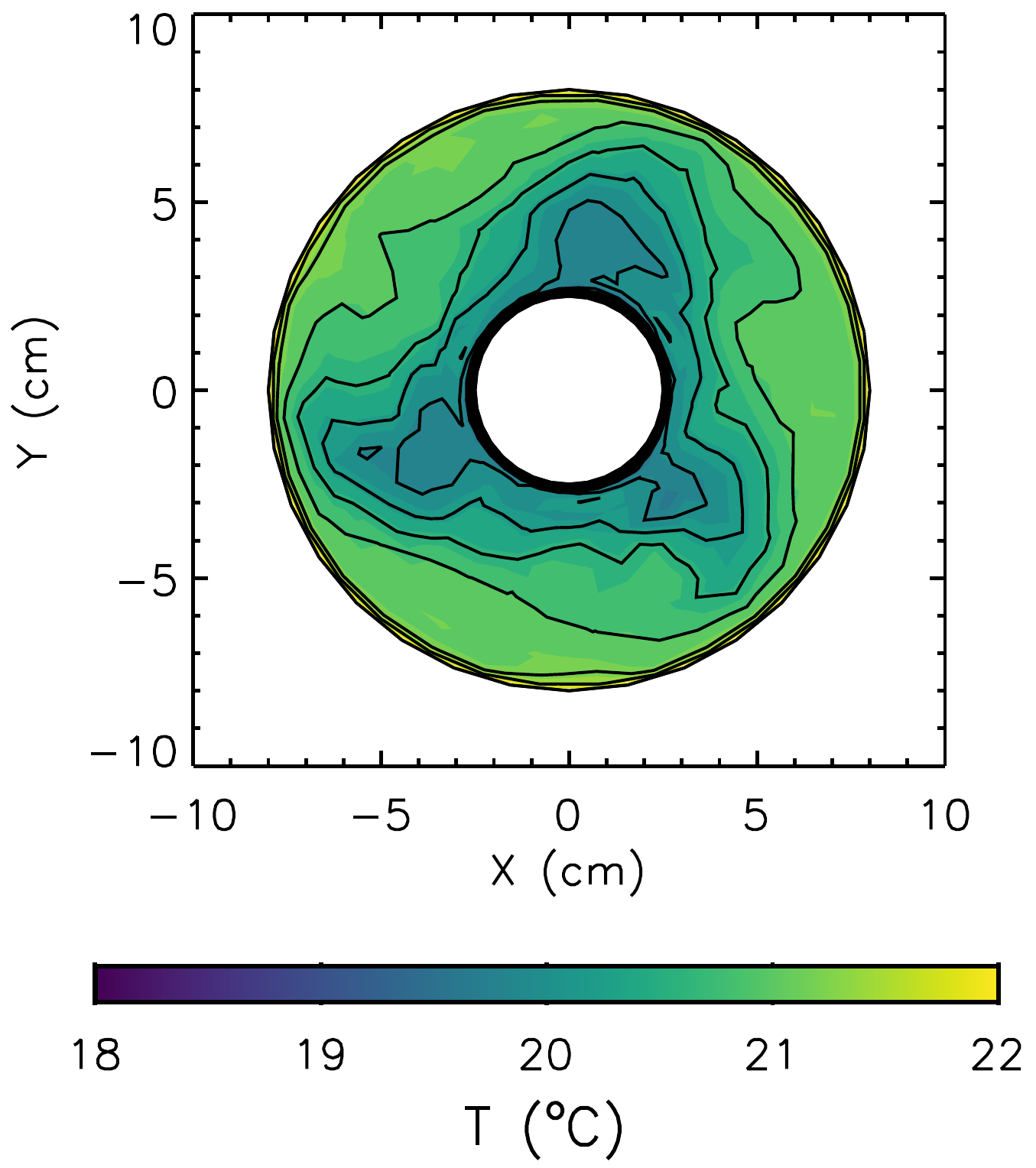}\,
	\includegraphics[height=0.38\textwidth,clip,viewport=2 222 295 410]{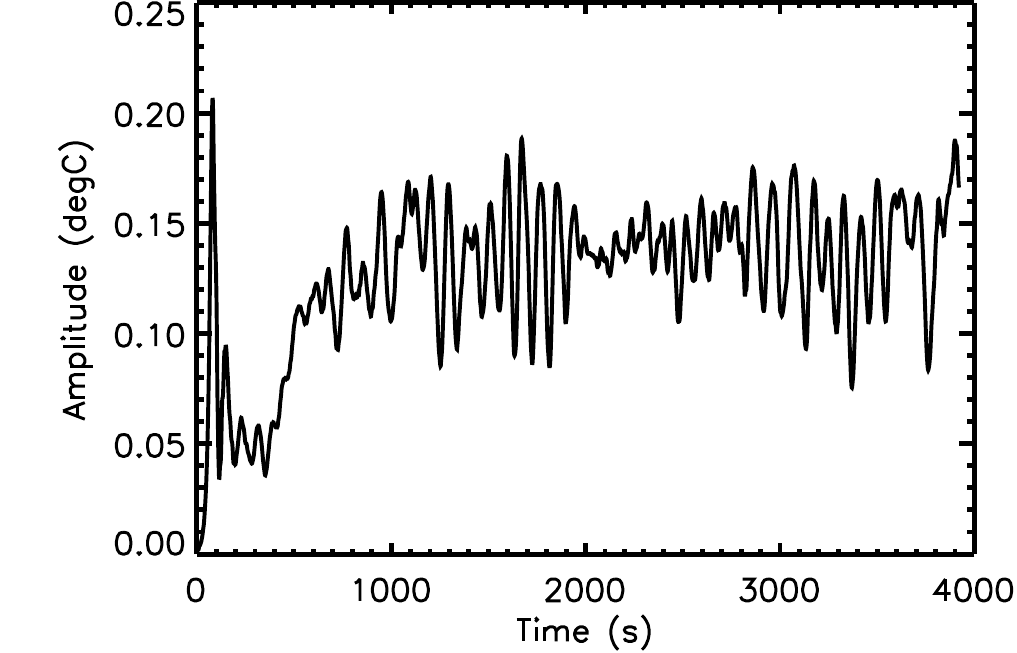}
	\caption{General flow appearance for the setup described in Sect.~\ref{sec:annulus}. Top: Horizontal snapshot through the temperature field at $z=5.61$\,cm after 2260\,s of simulation (contours every 0.2\,$^{\circ}$C). Bottom: Time series of the dominant wavenumber-3 mode amplitude for a 3960\,s simulation. The dominant mode amplitude is calculated by taking a Fourier transform over an azimuthal ring at mid-radius / mid-height each second during the run. The form of this time series is characteristic of chaotic annulus flow.}
	\label{fig:flow-appearance}
  \end{center}
\end{figure}

The mathematical model used to simulate the rotating annulus experiment shown in Fig.~\ref{fig:annulus-schem} is the Met Office / Oxford Rotating Annulus Laboratory Simulation (MORALS) \citep{1976Farnell,1985HignettB,2000Read}. The model is well established as a quanitatively accurate model of annulus flow in regular and weakly chaotic flow regimes; its details are given in the Appendix. The annulus setup is the ``small annulus'' configuration used by \citet{1985HignettB}. The model configuration is essentially the same as \citetalias{2019YoungQ}, and the flow simulated is also taken from that paper: rotation rate $\Omega=1$\,rad\,s$^{-1}$ and temperature difference between the cylinders of $\Delta T=4$\,degC. The dimension of the model is $N=24192$. With this setup and model resolution the general flow behaviour is shown in Fig.~\ref{fig:flow-appearance}, and the simulation displays chaotic dynamics. Table~\ref{tab:params} lists the annulus and MORALS parameters.

\section{Gradient descent of indeterminism}
\label{sec:gd}

Consider a sequence of states of a dynamical system $x_i$ valid at times $t_i$, $i=0,\ldots,w$, where $w$ is the \emph{window width}, and a model $f$ that maps the state $x_i$ forward in time from $t=t_i$ to $t_{i+1}$. Let each state have dimension $N$, hence the complete sequence defines a single point in $\field{R}^{N(w+1)}$. If
\begin{linenomath*}\begin{equation}
	x_{i+1}=f(x_i)\qquad i=0,\ldots,w-1
	\label{eq:traj}
\end{equation}\end{linenomath*}
then the sequence is a \textit{trajectory} of the model $f$, otherwise it is a \textit{$\delta$-pseudo-orbit} such that $|x_{i+1}-f(x_i)|<\delta\,\forall\,i$ \citep{1975Bowen}. The \textit{mismatch} between consecutive states is
\begin{linenomath*}\begin{equation}
	\delta x_i=x_{i+1}-f(x_i)
	\label{eq:mismatch}
\end{equation}\end{linenomath*}
The distance between the sequence and a model trajectory can be quantified using a scalar, the \textit{mean squared indeterminism} (or just the \textit{indeterminism}), which is the mean of the squared mismatches over the whole sequence:
\begin{linenomath*}\begin{equation}
	I = \frac{1}{w}\dsum_{i=0}^{w-1}\Vert \delta x_i\Vert^2\equiv\frac{1}{w}\dsum_{i=0}^{w-1}\Vert x_{i+1}-f(x_i)\Vert^2
	\label{eq:msi}
\end{equation}\end{linenomath*}
where $\Vert\cdot\Vert^2$ is the squared Euclidean norm. \textit{Gradient descent of indeterminism} solves the differential equation
\begin{linenomath*}\begin{equation}
	\frac{d{\bf x}}{d\tau}=-\frac{\partial I}{\partial\bf x}
	\label{eq:gd-diffeq}
\end{equation}\end{linenomath*}
where ${\bf x}=(x_0,x_1,\ldots,x_w)$, and ${\bf x}(\tau=0)={\bf s}$ is the initial sequence (raw observations or an analysis, perhaps). This equation defines how to change the sequence $\bf x$ in such a way that the indeterminism falls most quickly, relaxing the sequence of states onto the attractor of the model $f$. $I$ is a mathematical construct used to guide the gradient descent algorithm and to measure its progress; in general it does not have a physical interpretation. $\tau$ is called the \textit{descent time}, after \citet{2002Ridout}, and an intuitive graphical representation of the mechanism is shown in \citet[][Fig.~1]{2004JuddB}.

The algorithm itself is an iterative process. Denote state $i$ in the sequence after $h$ iterations by $x_{i,h}$. Each iteration is a two-step process. First, use the model to compute the forecast image $f(x_{i,h})$ and hence the mismatches $\delta x_i$ for each $i=0,\ldots,w-1$. Second, update the sequence using a discretization of Eq.~(\ref{eq:gd-diffeq}) \citep[Eq.~3]{2009Stemler}:
\begin{equation}
	x_{i,h+1} = x_{i,h}-\frac{2\,\Delta\tau}{w}\times \left\{\begin{array}{rcll}
	& - & \mathcal{A}(x_{0,h})\delta x_{0,h} & i=0\\
	& & & \\
	\delta x_{i-1,h} & - & \mathcal{A}(x_{i,h})\delta x_{i,h} & 1\leq i\leq w-1\\
	& & & \\
	\delta x_{w-1,h} & & & i=w
	\end{array}\right.
   \label{eq:gdf-update}
\end{equation}
where $\mathcal{A}(x_i)$ is the adjoint operator $df(x_i)^{\tt T}$ of $f$, and $\Delta\tau$ is a step length in $\field{R}^{N(w+1)}$. One can see from this definition that the change in each state is influenced by information propagated from earlier times via $\delta x_{i-1,h}$ and from later times via $\delta x_{i,h}$ mapped backwards in time by the adjoint operator. At the ends of the sequence information is propagated in one direction only, so the quality of the final sequence is expected to be poorer at the ends \citep[Fig.~3]{2002Ridout}. $h$ is then incremented by one and the procedure is repeated. The indeterminism can only reach zero in the asymptotic limit as $h\to\infty$, and then only in the perfect model scenario (see below), so in practice the algorithm is stopped when the indeterminism falls below a pre-defined minimum $\epsilon$, or manually after a certain number of iterations.

\subsection{Application to the MORALS perfect model scenario}

The perfect model scenario (PMS) provides a useful framework for exploring gradient descent and shadowing in complex systems. It allow many aspects of the experiments to be controlled, and provides a ``best case'' comparison for future results using laboratory data. The PMS simply means that the model $f$ and the system it is modelling, $\tilde{f}$, are equivalent. In this work we set up the PMS by taking both model and system as MORALS simulations with the same parameters as \citetalias{2019YoungQ}.

This scenario offers a number of advantages, the most useful of which is that the true state is known exactly and thus explicit comparisons of forecasts with truth can be made. We can define one model run as the true trajectory of the system, and generate artificial observations from that using a known noise model. The PMS offers the greatest amount of control over the experimental configuration, which allows us to study the properties of the system and algorithm in isolation.

In the PMS, gradient descent will converge to a model trajectory under certain conditions \citep[Proposition 2]{2002Ridout}. \citet{2004JuddB} proved the surprising result that gradient descent still works for incomplete or even wholly unrealistic adjoint operators. In particular, they showed that setting $\mathcal{A} = \lambda {\bf I}$, where $\lambda$ is a scalar and $\bf I$ is the identity matrix, is sufficient. They call this \textit{gradient-free descent}, as the method can then be used without knowledge of the gradient of the operator $f$. The conceptual change compared with using the true adjoint is that the algorithm now moves the state in a direction of decreasing indeterminism, but not in the direction of \textit{steepest} descent. We use gradient-free descent here, referring to $\lambda {\bf I}$ as the \textit{$\lambda$-adjoint}, because there is currently no adjoint model available for MORALS.

Using this construction we can also test the method with respect to the shadowing properties of candidate trajectories produced from its output. We know \textit{a priori} that the true shadowing time is the whole observation sequence because we already know a model trajectory exists that shadows the observations: the true trajectory the observations were generated from. We expect the candidate shadowing times to be sensitive to how close they begin from that true trajectory, although this does not always hold in practice.

The specifics of how the method was applied to MORALS are detailed in the Appendix, as it differs from the general outline above primarily in matters of notation. In what follows we denote, in general, a state of the annulus model defined on the MORALS grid by the vector $\bf x$ with $\dim({\bf x})=N$. We use $\mathcal{X}$ to denote, in general, a sequence of such model states:
\begin{linenomath*}\begin{equation}
	\mathcal{X}\equiv({\bf x}_0, {\bf x}_1, \ldots, {\bf x}_i, \ldots, {\bf x}_w)
\end{equation}\end{linenomath*}
for a sequence of length $w+1$, and hence $\dim(\mathcal{X})=N(w+1)$, where the states are valid at times $t_i$. We denote by ${\bf x}_{i,h}$ a single state in the sequence (or \textit{shadow analysis}) after $h$ completed iterations of the gradient descent algorithm. $\mathcal{X}_h$ is the whole sequence of $w+1$ states (the \textit{sequence of shadow analyses}) after $h$ gradient descent iterations, and hence $\mathcal{X}_0$ denotes the sequence of observations. We denote the true sequence by $\hat{\mathcal{X}}$, a single state in that true sequence by $\hat{\bf x}_i$, and a generic true state by $\hat{\bf x}$.

\section{A demonstration of gradient descent using the annulus}
\label{sec:results}

We now present a demonstration of gradient descent using the annulus, followed by an analysis of the $\lambda$ parameter (the parameter that multiplies the identity matrix in gradient-free descent). In the next section we calculate shadowing times for candidate trajectories using the shadow analyses to generate initial conditions. The gradient descent was started from an observational sequence of 65 states separated by 5\,s, and ran with $\lambda=0.5$ for 500 iterations. Other parameters are listed in the Appendix. The sequence is long enough to cover about four periods of the longest timescale associated with the flow. Observations were generated by adding normally-distributed random numbers to a sequence of true states calculated using the model. The random numbers represent observational error of standard deviation $\sigma=1/3$ of the natural variability of the model at each grid point, denoted by $\bf r$. The method used to calculate $\bf r$ is detailed in the Appendix.

\begin{figure*}[p]
  \centering
	\includegraphics[width=0.495\textwidth,clip,viewport=33 25 810 461]{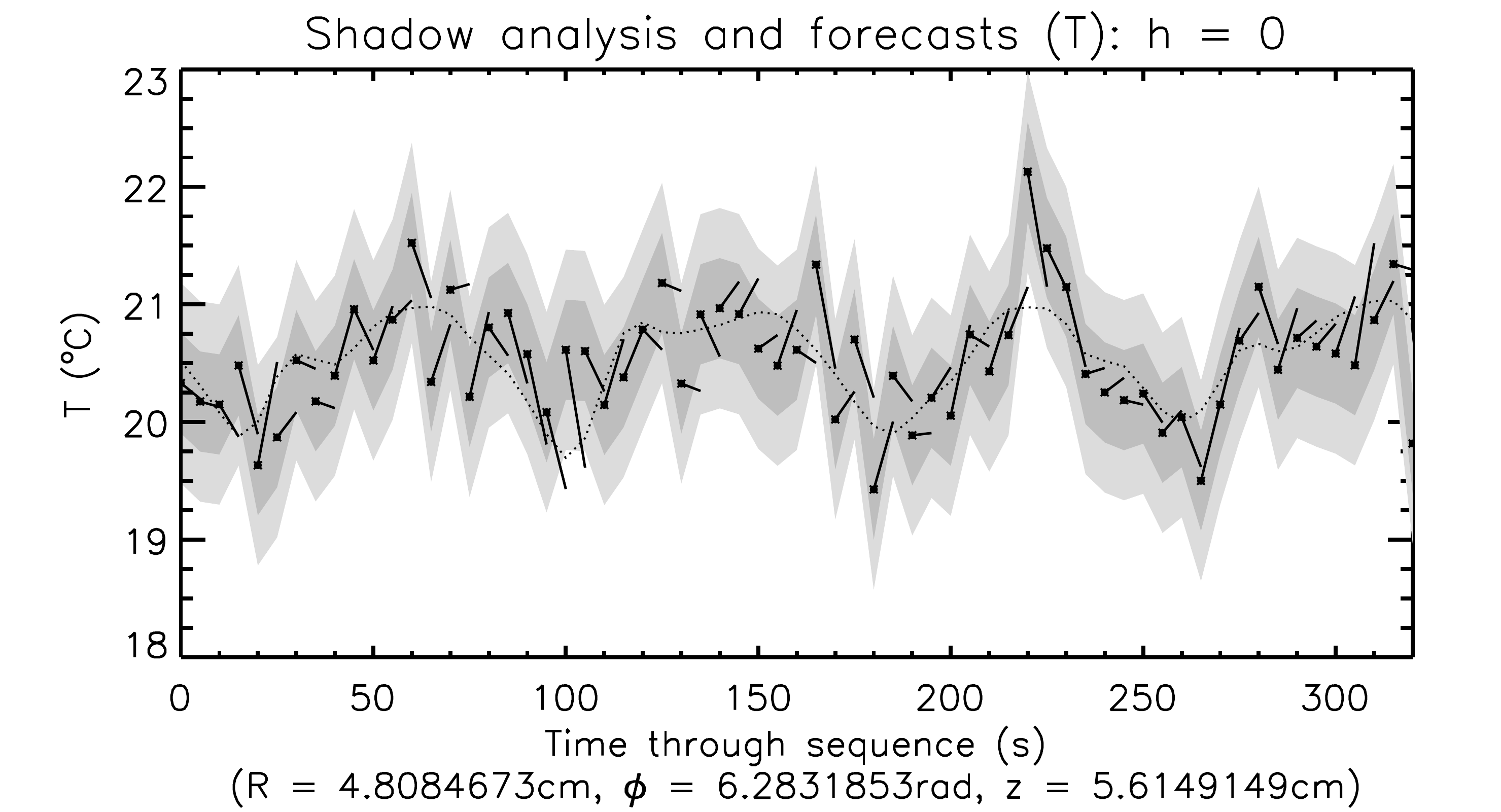}
	\includegraphics[width=0.495\textwidth,clip,viewport=33 25 810 461]{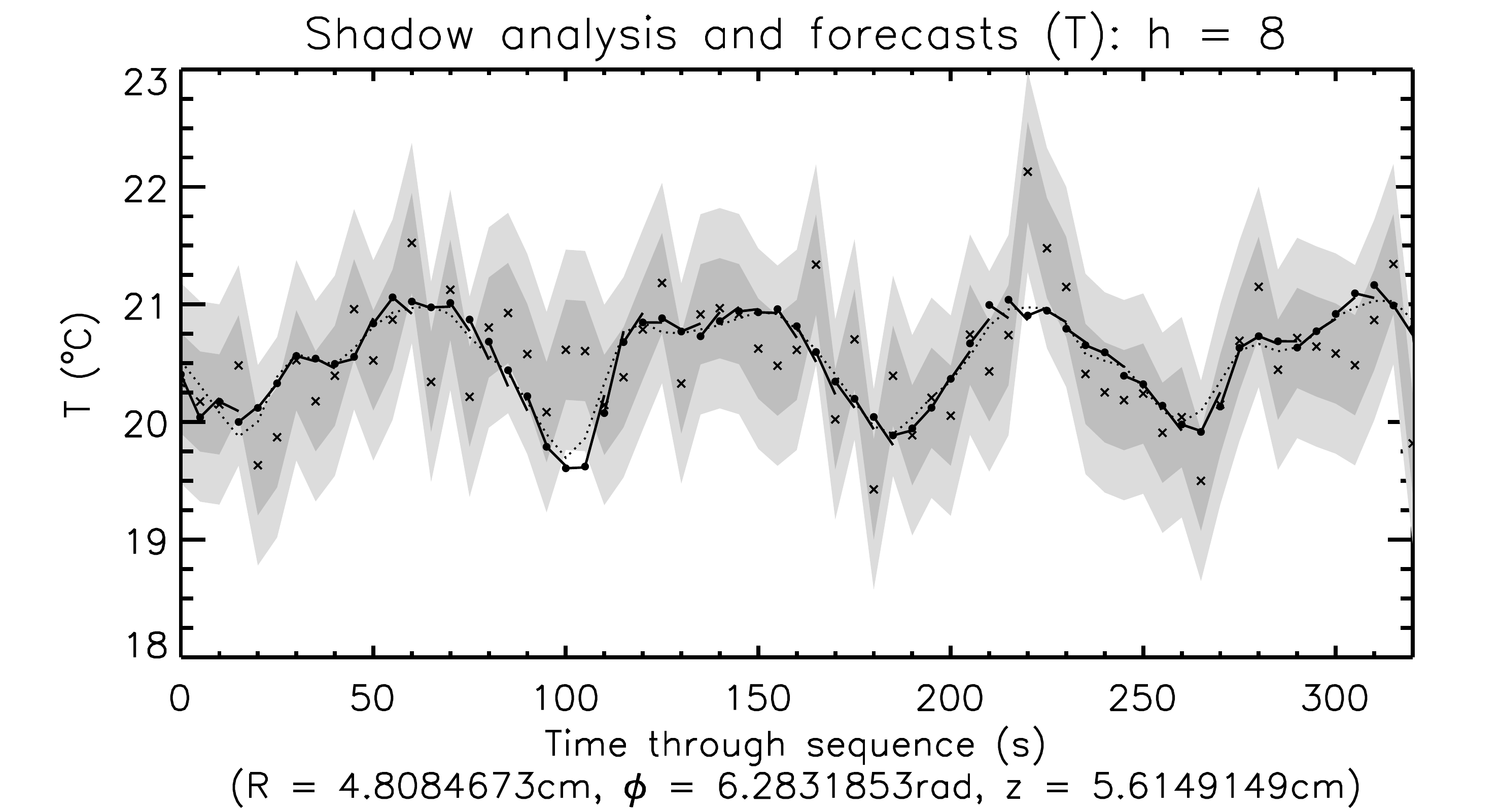}\\
	\includegraphics[width=0.495\textwidth,clip,viewport=33 25 810 461]{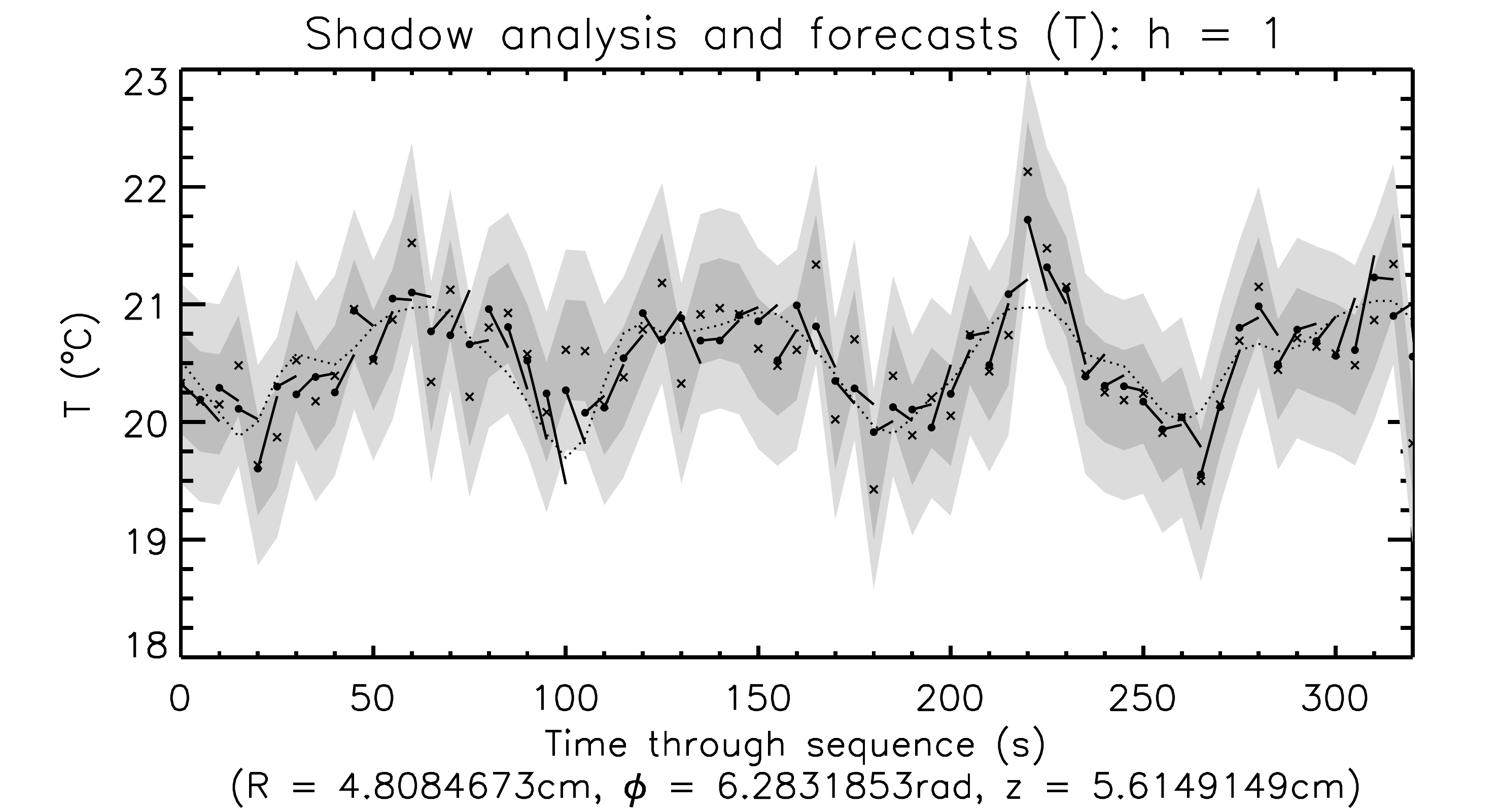}
	\includegraphics[width=0.495\textwidth,clip,viewport=33 25 810 461]{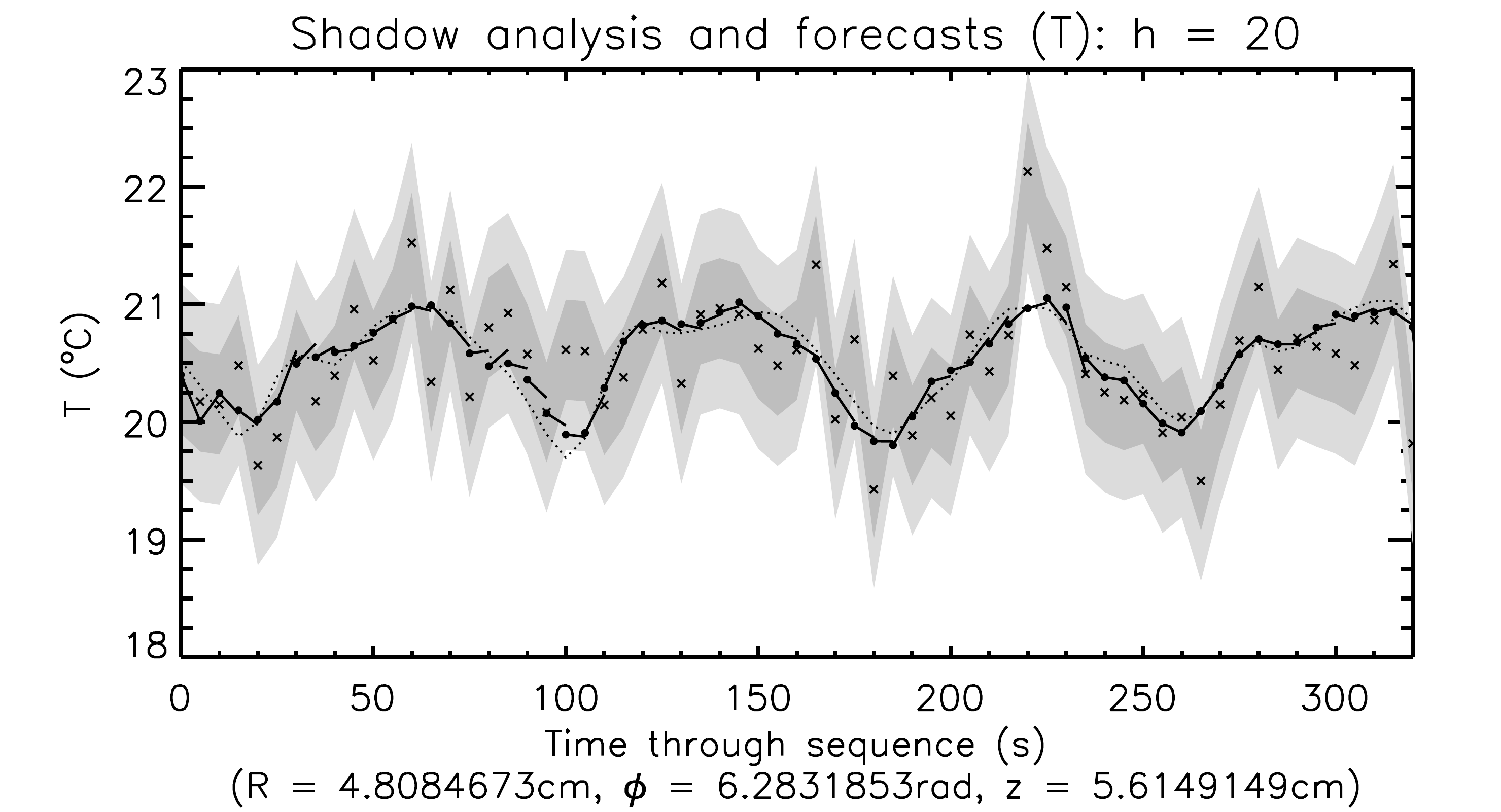}\\
	\includegraphics[width=0.495\textwidth,clip,viewport=33 25 810 461]{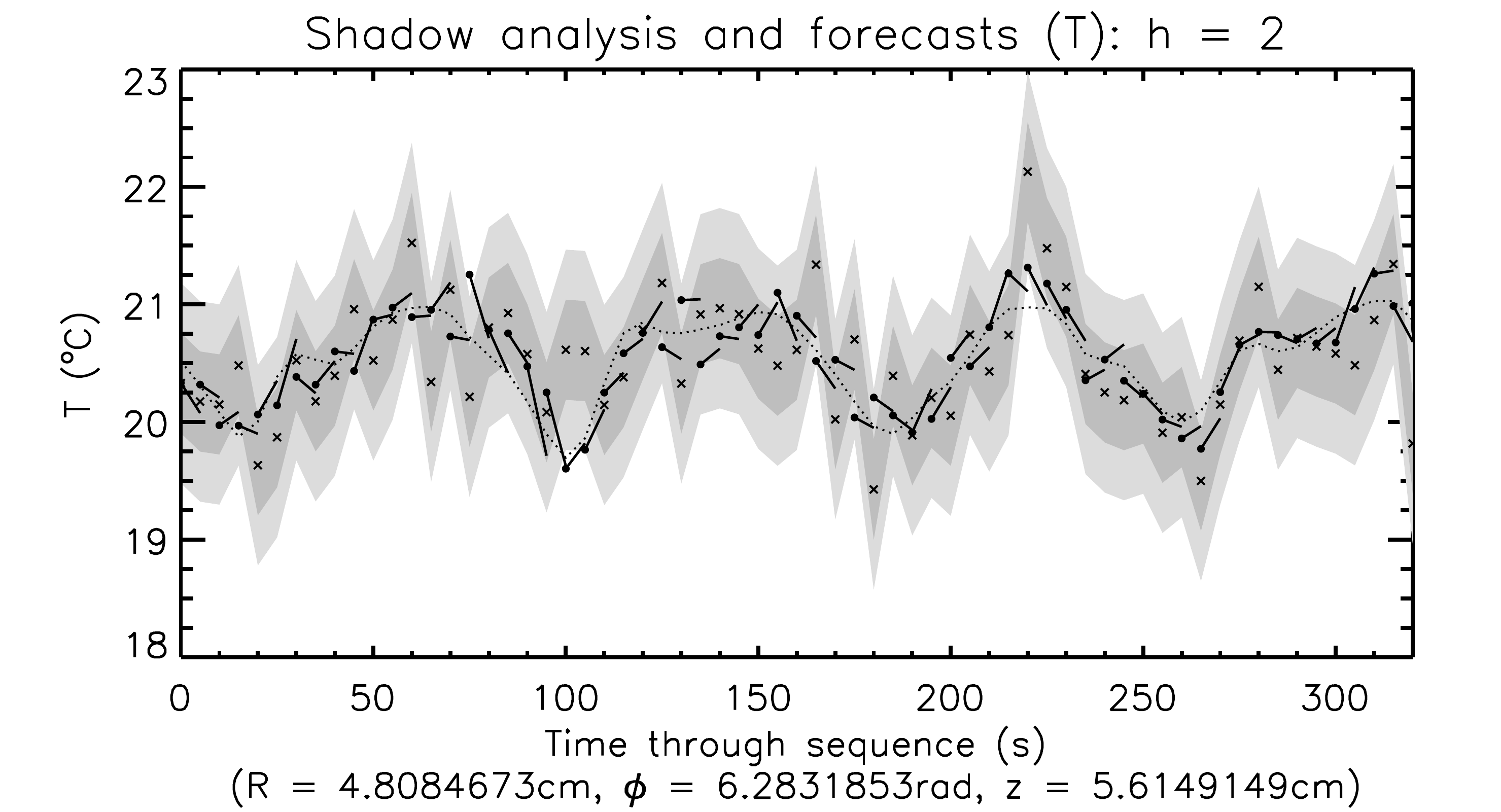}
	\includegraphics[width=0.495\textwidth,clip,viewport=33 25 810 461]{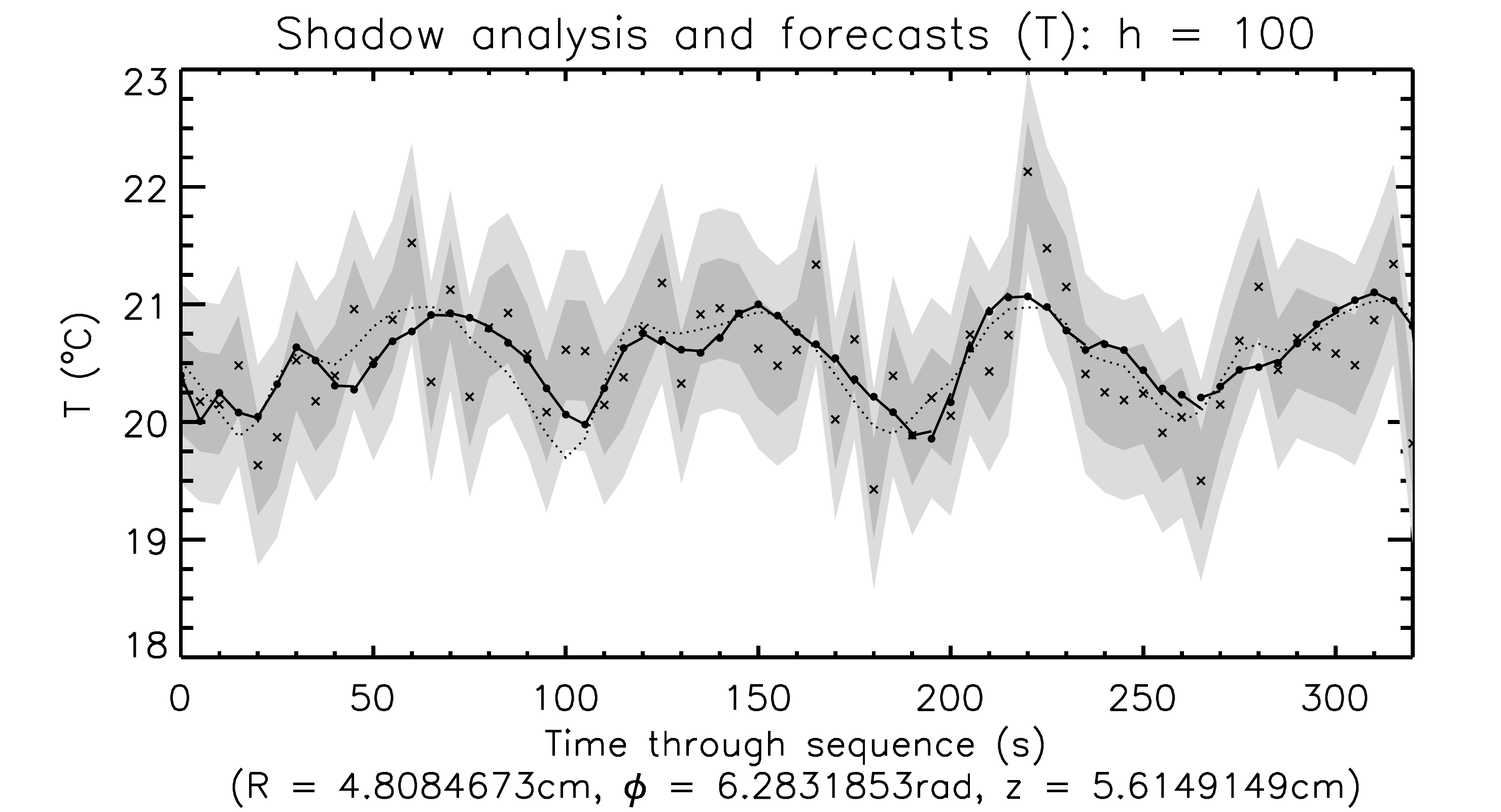}\\
	\includegraphics[width=0.495\textwidth,clip,viewport=33 25 810 461]{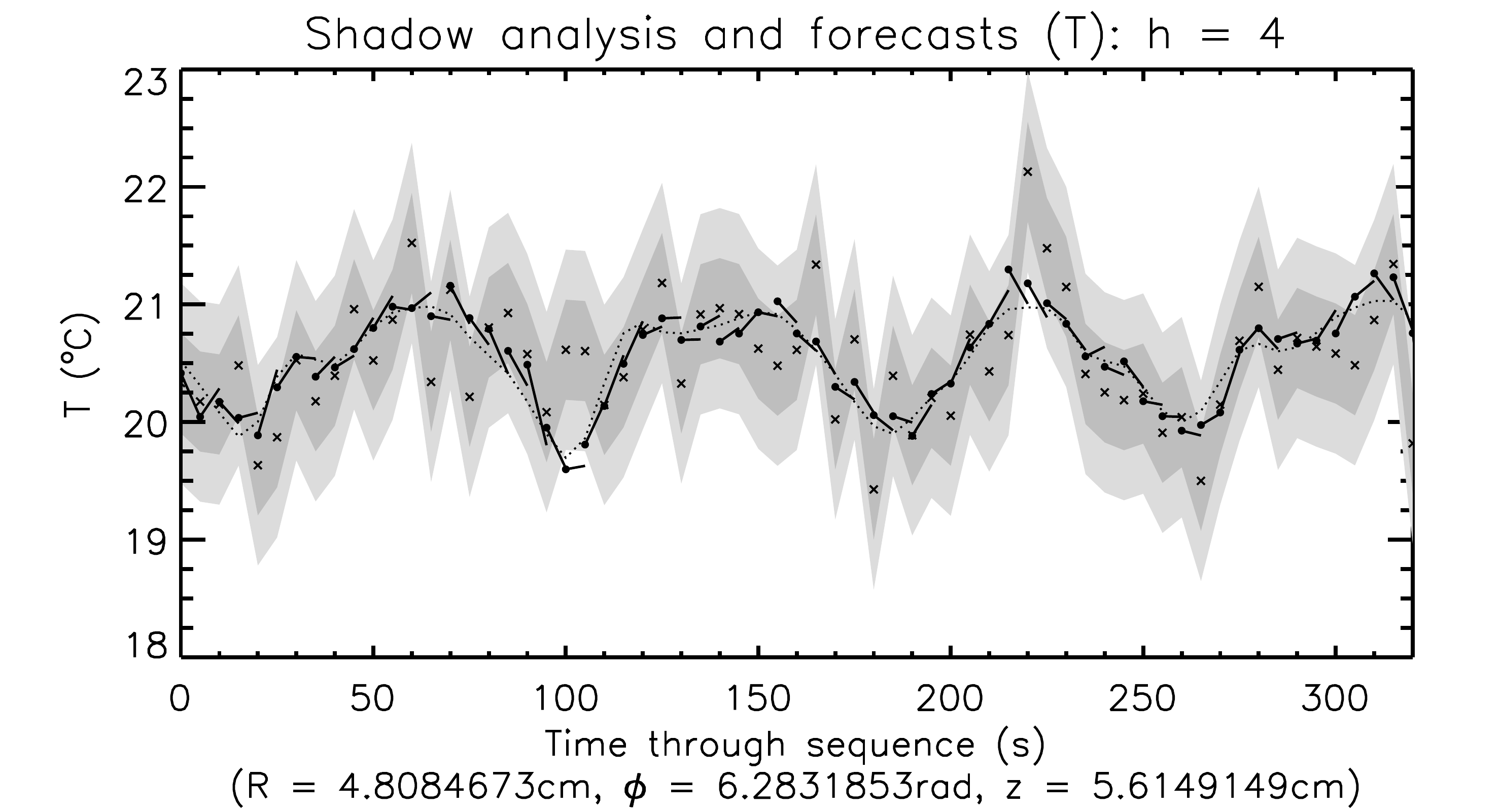}
	\includegraphics[width=0.495\textwidth,clip,viewport=33 25 810 461]{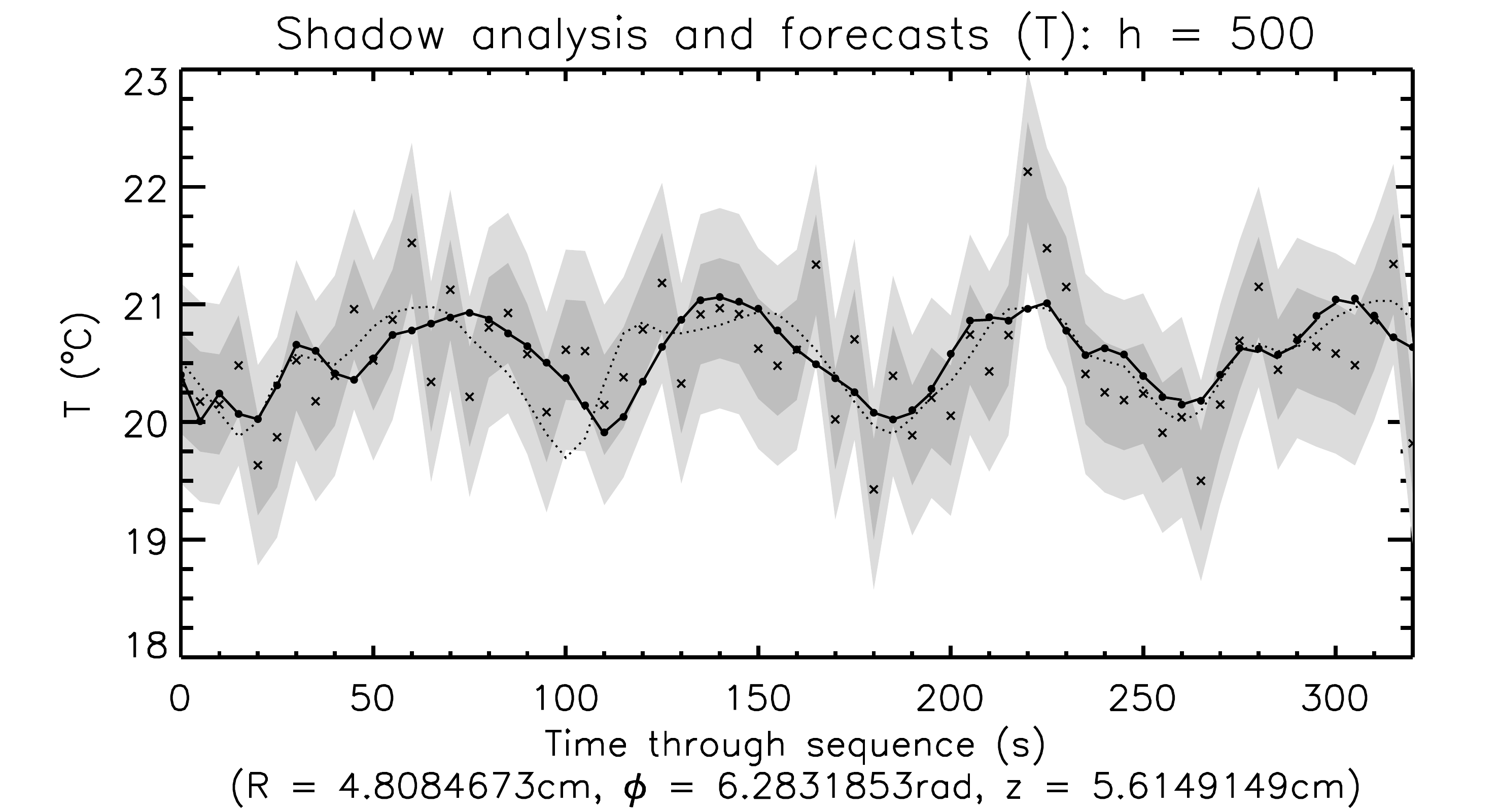}
  \caption{Progression of the gradient descent at a single point. Each panel shows the temperature at a single grid point in the model ($R=4.81$\,cm, $\theta=6.18$\,rad, $z=5.61$\,cm) as a function of time over the gradient descent window. The sequence is shown at the following iterations: $h=0$, 1, 2, 4, 8, 20, 100, and 500. Dots show the shadow analyses, crosses show the observations, solid lines join each shadow analysis at time $t_i$ with its forecast image at time $t_{i+1}$, grey shaded areas show the range spanned by the observation $\pm 1\sigma$ error (darker shade) and $\pm 2\sigma$ error (lighter shade), and the dotted line is the truth (unknown to the gradient descent algorithm, but included for comparison with the final shadow analysis sequence and the original observations).}
  \label{fig:descent-single}
\end{figure*}

\begin{figure*}[p]
\vspace*{-1.5cm}
  \centering
	\includegraphics[width=\textwidth,clip,viewport=30 201 1277 546]{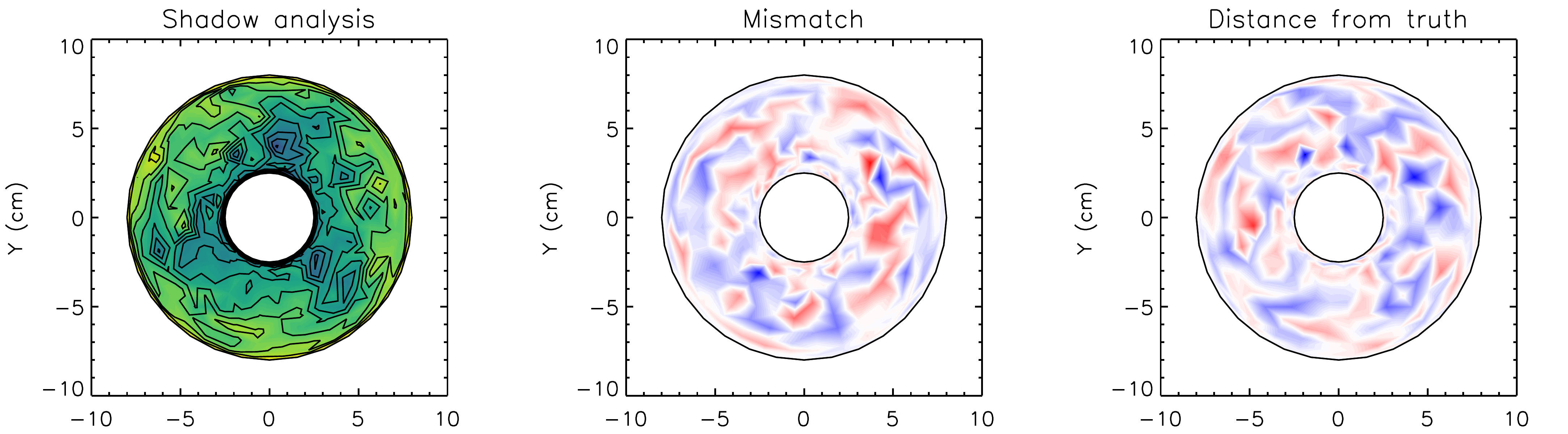}\\
	\includegraphics[width=\textwidth,clip,viewport=30 201 1277 518]{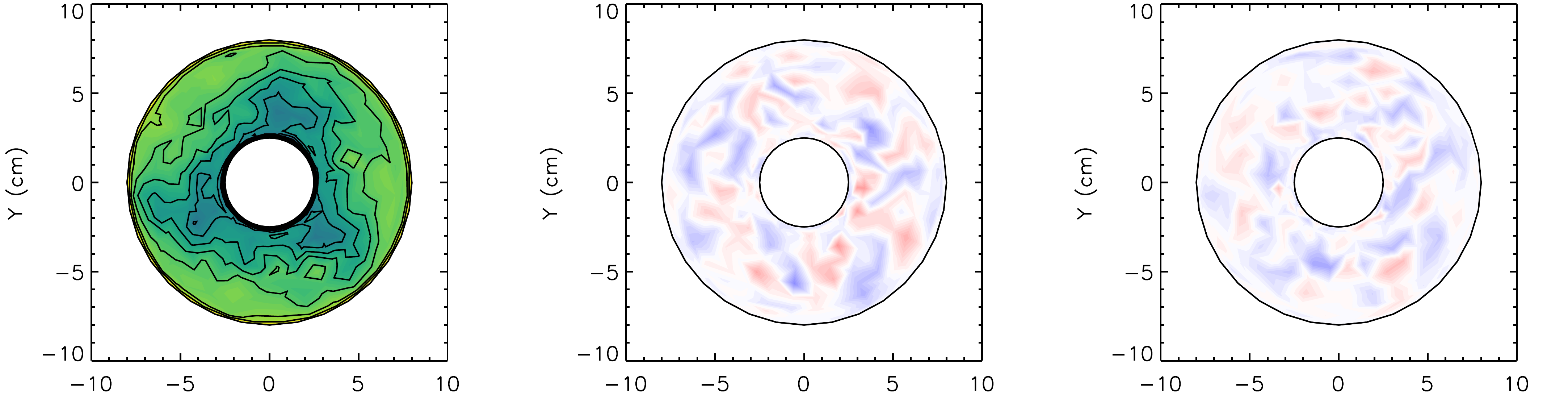}\\
	\includegraphics[width=\textwidth,clip,viewport=30 201 1277 518]{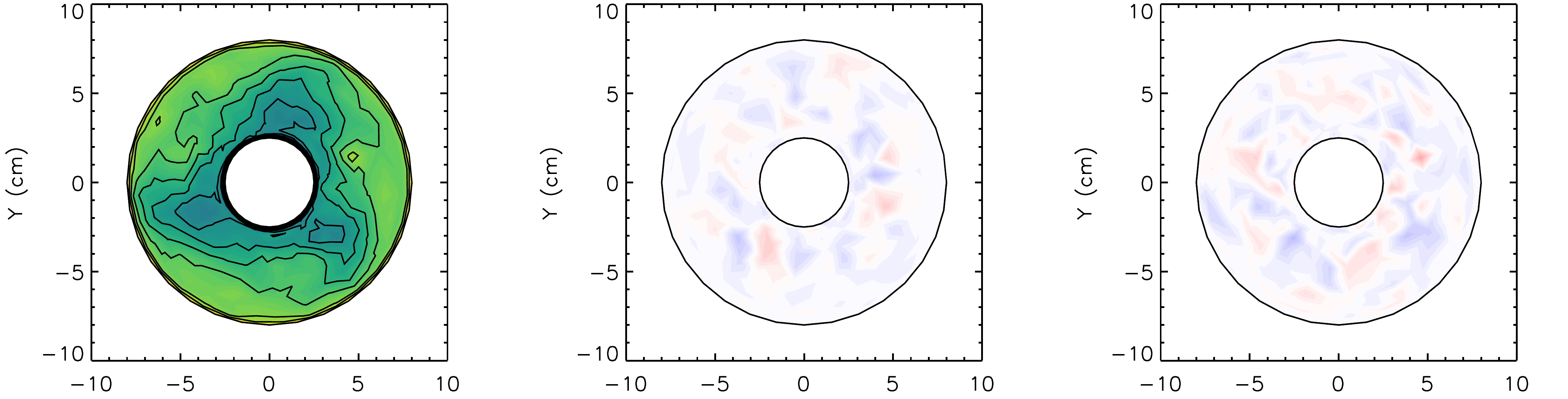}\\
	\includegraphics[width=\textwidth,clip,viewport=30 201 1277 518]{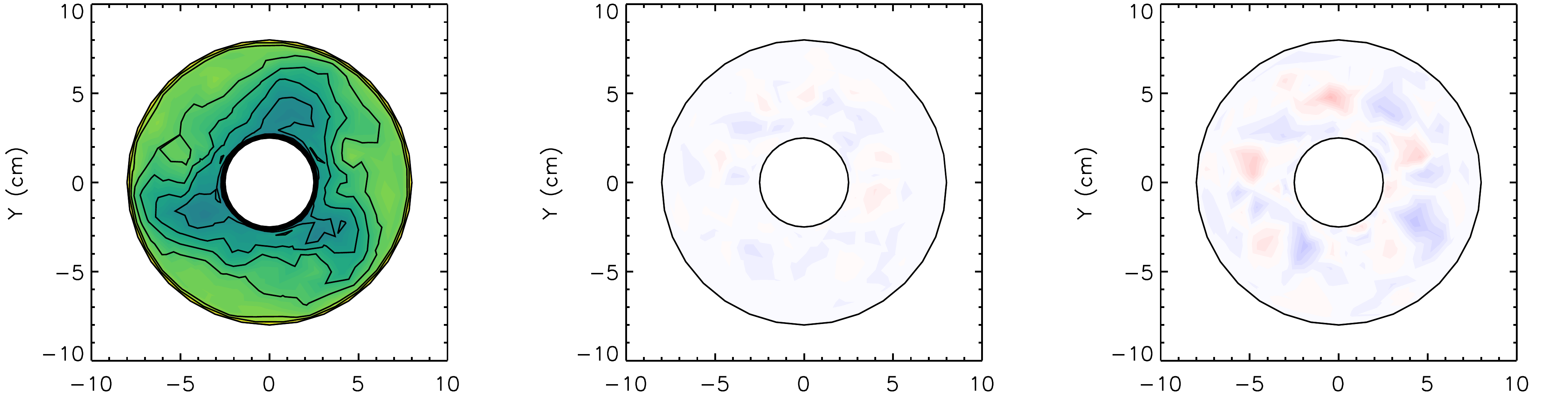}\\
	\includegraphics[width=\textwidth,clip,viewport=30 170 1277 518]{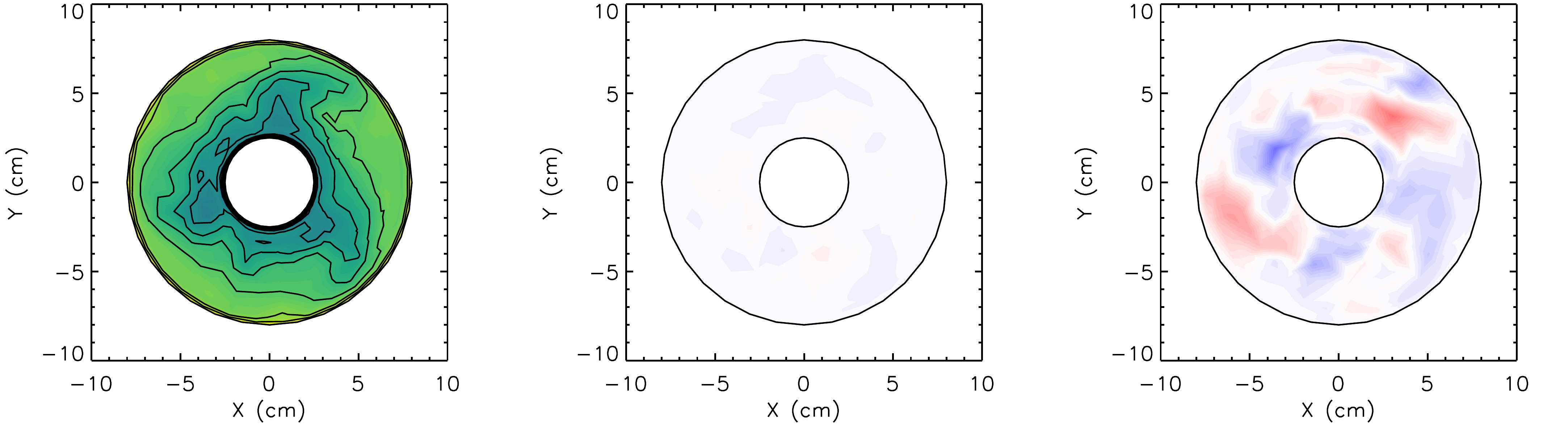}\\
	\includegraphics[width=\textwidth,clip,viewport=30 80 1277 160]{hires3_h0500_i0032_single_horiz_T.pdf}
  \caption{Progression of the gradient descent shown as a horizontal slice through the temperature field at $z=5.61$\,cm and $t=t_0+160$\,s ($i=32$, the mid-point in the sequence). The gradient descent is shown at the following iterations: $h=0$, 2, 4, 8, and 500 (top to bottom). Three quantities are shown: (left) the shadow analysis ${\bf x}_{i,h}$ after $h$ gradient descent steps, (middle) the forecast mismatch $\delta{\bf x}_{i,h}$ corresponding to that shadow analysis, and (right) the difference between the shadow analysis and the truth ${\bf x}_{i,h}-\hat{\bf x}_{i}$. The left panel of Fig.~\ref{fig:flow-appearance} shows the truth for this particular run.}
  \label{fig:descent-slices}
\end{figure*}

\subsection{Visual demonstration of the gradient descent}

In Figs~\ref{fig:descent-single} and \ref{fig:descent-slices} we show how $\mathcal{X}_h$ changes during the gradient descent. Figure~\ref{fig:descent-single} shows the progression of the time series at a single grid point, while Fig.~\ref{fig:descent-slices} shows a horizontal section through the annulus temperature field near mid-height. These quantities are shown at several steps during the gradient descent. The PMS allows us to see directly how well gradient descent recovers the initial true trajectory $\hat{\mathcal{X}}$, while using information only from the model and observations.

\begin{figure}[tb]
  \centering
	\includegraphics[width=0.495\columnwidth,clip,viewport=30 8 570 427]{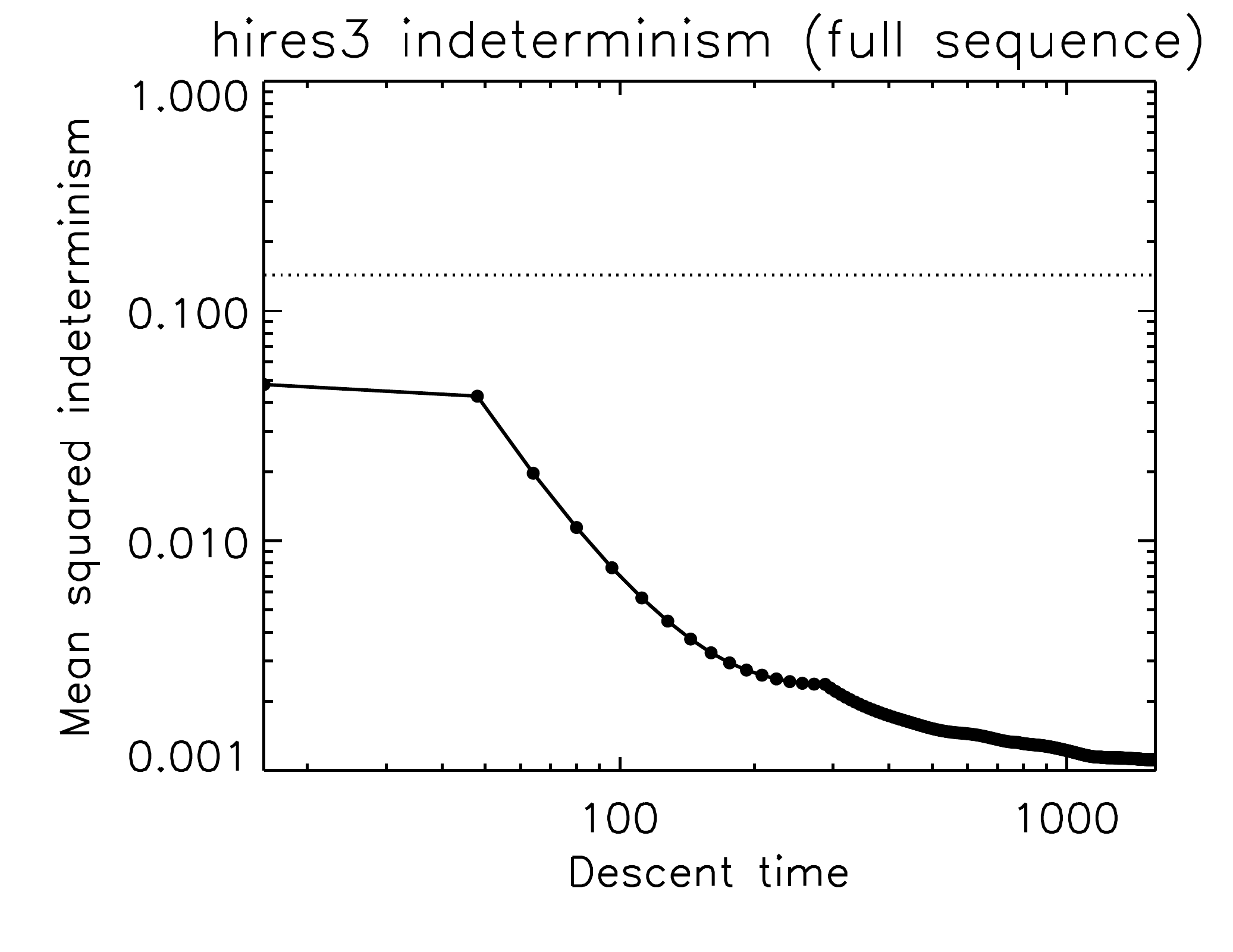}\,
	\includegraphics[width=0.495\columnwidth,clip,viewport=30 8 570 427]{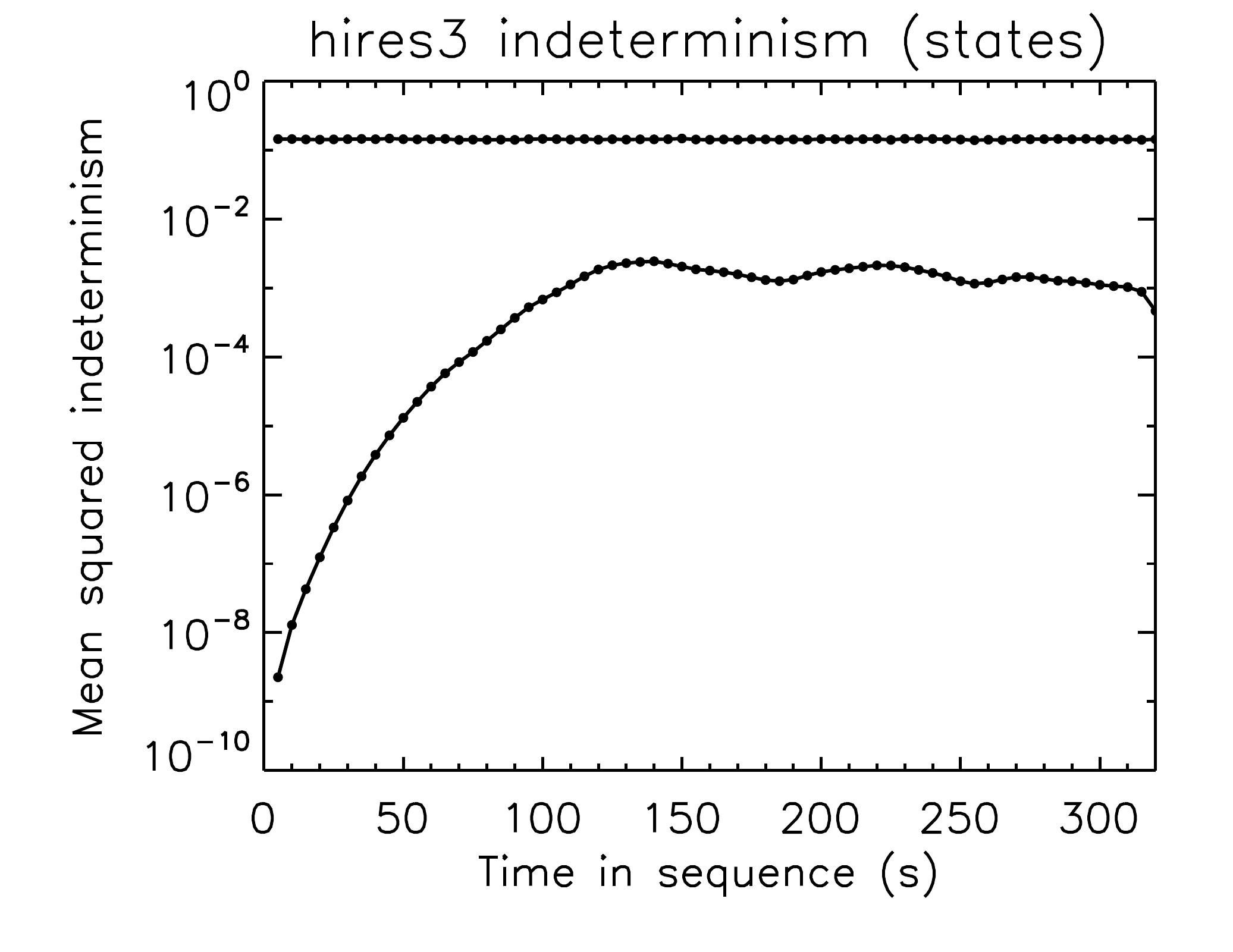}
	\caption{Mean squared indeterminism for the gradient descent in Figs~\ref{fig:descent-single}--\ref{fig:descent-slices}. The left panel shows the total indeterminism  $I(\mathcal{X}_h)$ (Eq.~\ref{eq:msi-morals-2}) as a function of descent time $\tau$. Each iteration is represented by a dot, and the horizontal dotted line shows the value at $h=0$. The right panel shows, as a function of position in $\mathcal{X}_h$, the mean squared indeterminism for each state $I({\bf x}_{i,h})$ (Eq.~\ref{eq:indeterminism-state1}) at the start of the gradient descent ($h=0$, upper line) and at the end of the gradient descent ($h=500$, lower line).}
	\label{fig:hires3-indeterminism}
\end{figure}

It is easiest to visualise the progress of the gradient descent using the time series in Fig.~\ref{fig:descent-single}. At the start of the gradient descent ($h=0$) the states and their forecast images (black dots and solid lines) do not join up at all; $\mathcal{X}_0$ is far from a model trajectory. During the first several steps of the gradient descent $\mathcal{X}_h$ falls quickly towards a model trajectory. The top panel of Fig.~\ref{fig:hires3-indeterminism} shows $I(\mathcal{X}_h)$ as the gradient descent progresses, as a function of the descent time
\begin{linenomath*}\begin{equation}
  \tau(h)=\dsum_{j=0}^{h-1}\Delta\tau_j
\end{equation}\end{linenomath*}
Indeterminism falls off approximately as a power law during the first few gradient descent steps. By $h=8$ the time series in Fig.~\ref{fig:descent-single} is close enough to the trajectory to require closer inspection to confirm the time series is not a trajectory, but a pseudo-orbit.

Figure~\ref{fig:descent-slices} complements the time series in Fig.~\ref{fig:descent-single} by showing how the gradient descent progresses over a whole horizontal section. Three different quantities are shown in Fig.~\ref{fig:descent-slices} as horizontal sections: the shadow analysis ${\bf x}_{i,h}$, mismatch ${\bf x}_{i+1,h}-f({\bf x}_{i,h})$, and distance from truth ${\bf x}_{i,h}-\hat{{\bf x}}_i$, all for $i=32$, the mid-point in $\mathcal{X}_h$. First, from the middle of these panels we see that the most striking change over the gradient descent is the mismatch. Like the indeterminism in Fig.~\ref{fig:hires3-indeterminism}, this falls off very quickly during the gradient descent such that by $h=8$ the mismatch is some two orders of magnitude smaller than at $h=0$. This decrease is reflected in the colour scale in that figure. Second, the shadow analysis ${\bf x}_{i,h}$ on the left starts off quite noisy and by the bottom of the figure much of the noise has been smoothed out. Finally, on the right the distance between the shadow analysis and the truth also falls rapidly, although not as fast as the indeterminism. Unlike the fall in indeterminism, however, the distance from the truth does not fall off monotonically but begins to increase again by $h=20$; we shall examine this in more detail below.

Figures~\ref{fig:descent-single} and \ref{fig:descent-slices} demonstrate the ability of gradient descent to recover a sequence of states much closer to a model trajectory than the original $\mathcal{X}_0$ in less than ten iterations. While $\hat{\mathcal{X}}$ is included in Fig.~\ref{fig:descent-single}, the gradient descent algorithm has no information about the truth at all, only the observations and the model. 

To check the reproducibility of these results, we ran nine additional gradient descents using exactly the same setup except with different random numbers used to generate the observations. All ten cases produced very similar results, with no outliers. We also ran two additional gradient descents started from different points in the model's state space, but with an otherwise identical setup. Again, the results were very similar. This reinforces the conclusion that the initial part of the gradient descent primarily removes noise from the observations, and the latter part converges towards a trajectory; in each case the noise statistics are the same, but the results only diverge once indeterminism has fallen by two orders of magnitude. 

Figure~\ref{fig:hires3-indeterminism} shows that, overall, $\mathcal{X}_{500}$ is two orders of magnitude closer to a trajectory than the original $\mathcal{X}_0$, both from the full indeterminism in the top panel of that figure, and by comparing the state-wise indeterminism between $\mathcal{X}_0$ and $\mathcal{X}_{500}$ in the lower panels. At the start of the gradient descent the state indeterminism $I({\bf x}_{i,h})$ (Fig.~\ref{fig:hires3-indeterminism}, bottom) is approximately constant with position in the sequence. By construction, the initial expected squared distance from truth at each grid point is approximately $(1/3)^2\simeq0.11$, the variance of the observational error accounting for scaling by $\bf r$.

After 500 gradient descent steps, however, there is a clear structure to the variation of indeterminism with position in the sequence (Fig.~\ref{fig:hires3-indeterminism}, bottom, lower line). At the beginning of the sequence there is an approximately exponential growth in indeterminism. This can be explained by the choice of $\lambda$. As $\lambda<1$ more weight is assigned, in the update step in Eq.~(\ref{eq:gdf-update}), to $\delta{\bf x}_{i-1,h}$ compared with $\delta{\bf x}_{i,h}$. Hence more information is passed forwards compared with backwards in time, and so during the gradient descent any given state will tend to reduce the mismatch compared with the state before it faster than the mismatch compared with the state after it. Hence the mismatch at the start of the sequence will decrease the fastest. We shall see later how the value of $\lambda$ affects this.

\subsection{Correspondence between shadow analysis and truth}

\begin{figure}[tb]
  \centering
	\includegraphics[width=0.495\textwidth,clip,viewport=40 20 570 395]{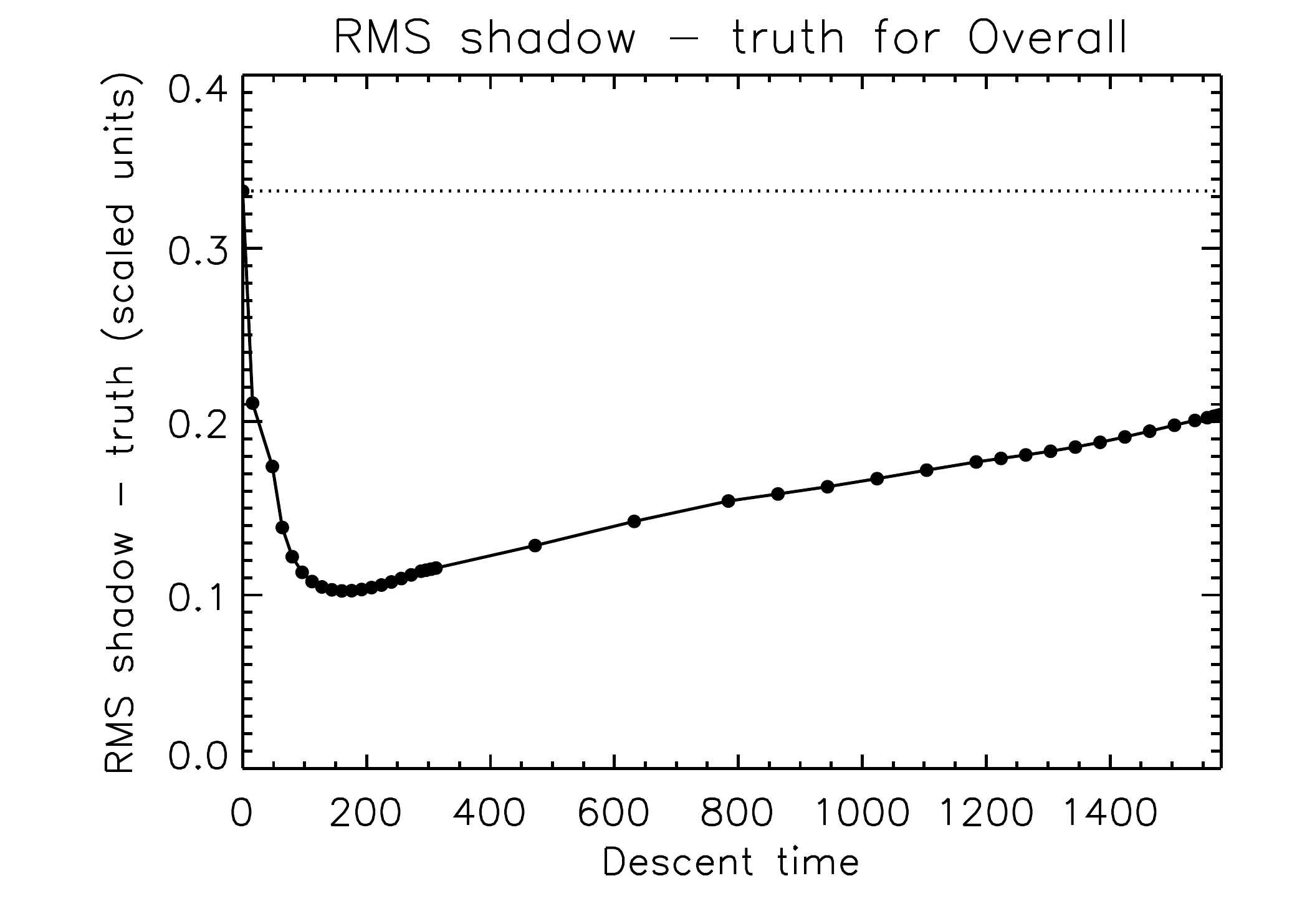}
	\caption{RMS Euclidean distance per grid point scaled by $\bf r$ between the sequence of shadow analyses and truth (Eq.~\ref{eq:truth-dist}) as a function of the descent time $\tau$. The distance is plotted for each iteration up to $h=20$ and thereafter for iteration numbers divisible by 20. The horizontal dotted line shows the value at $h=0$ for ease of comparison with later values.}
	\label{fig:truth-dist}
\end{figure}

We have demonstrated that gradient descent recovers a sequence of states close to a model trajectory, but our main reason for its use is that it often produces candidate states with longer shadowing times than other methods used to generate candidates. Above we noted that one trajectory guaranteed to shadow the observations is the original true trajectory, so the correspondence between our pseudo-orbit of shadow analyses and the true trajectory is important. Whether $\mathcal{X}_h$ is close to truth can be measured directly, as we are in the PMS. Figure \ref{fig:truth-dist} shows, as a function of descent time $\tau$, the RMS Euclidean distance per grid point between $\mathcal{X}_h$ and the truth $\hat{\mathcal{X}}$ (the distance between two points in $\field{R}^{N(w+1)}$) scaled by $\bf r$:
\begin{linenomath*}\begin{equation}
	D(h) = \left[\frac{1}{N(w+1)}\dsum_{i=0}^w\Vert({\bf x}_{i,h}-\hat{\bf x}_i)\circ{\bf r}^{-1}\Vert^2\right]^{1/2}
	\label{eq:truth-dist}
\end{equation}\end{linenomath*}
This distance is expressed per grid point so the values are easily comparable with the observational noise standard deviation $\sigma$. By construction, the initial distance from truth is approximately $\sigma=1/3$, as this distance just corresponds to the observational error. The distance from truth then falls off quickly during the first few iterations. After 10--15 iterations, however, the distance from the truth stops falling and begins to rise again, which it does so for the remainder of the 500 iterations. So while $I(\mathcal{X}_h)$ is monotonic during the gradient descent, $\mathcal{X}_h$ approaches $\hat{\mathcal{X}}$ but then moves away again. In this particular example the closest approach to $\hat{\mathcal{X}}$ is about one third of the original distance from it.

This can also be seen in the right panels of Fig.~\ref{fig:descent-slices}. The final three steps show the distance from truth increasing after $h=8$ (which corresponds to $\tau=144$, for orientation in Fig.~\ref{fig:truth-dist}), and in the time series in Fig.~\ref{fig:descent-single} we can see $\mathcal{X}_h$ move away from the truth by steps $h=20$, 100, and 500 ($\tau=312$, 864, and 1579.1875). When the mismatch at a single point in $\mathcal{X}_h$ is large compared with adjacent points, such as between $t=70$ and 100\,s at $h=20$ in Fig.~\ref{fig:descent-single}, this mismatch propagates along the sequence, introducing a phase error in the position of the baroclinic wave along $\mathcal{X}_h$ when compared to the values in $\hat{\mathcal{X}}$.

Insight from \citet{2009Stemler} goes some way to explaining this result. The primary reason appears to be the $\lambda$-adjoint approximation we are using. They performed a systematic comparison of different adjoint approximations using the \citet{1963Lorenz} system, measuring, among other things, indeterminism and distance from truth as a function of computational cost, which corresponds loosely to the number of iterations here. They found that, after about 60 gradient descent iterations using the $\lambda$-adjoint (equivalent cost to about 30 steps in our case, as we have 65 states in the sequence while they have 30), the distance from truth began to increase again and eventually diverged (their Fig.~9). They do note that the indeterminism also increases, which is possible in their experiments because they do not change $\Delta\tau$ during the gradient descent, while we do, but they note that the distance from truth begins to increase some time \textit{before} the indeterminism increases.

To explain this result, they note that ``the first iterations of the shadowing filter tend to remove effects of observational noise, and subsequent iterations achieve convergence to a trajectory''. In the first iterations a full adjoint doesn't provide much extra information for the gradient descent, but in the later stages it is vital. \citet{2008Judd} examined this more closely using both the PMS and IMS. He argued that (1) when $\mathcal{X}_h$ is altered in a way that moves perpendicular to the indeterminism contours the surfaces of constant indeterminism ``are well-behaved with smooth slow variations'', while (2) when moving at an angle there is a complex local variation in indeterminism close to $I=0$. He demonstrates this point using a full adjoint model in the context of the PMS (case 1) and IMS (case 2), but the same principle can be applied in the PMS to a comparison of the full adjoint (case 1) and $\lambda$-adjoint (case 2), in which indeterminism is decreased but not in the direction perpendicular to the indeterminism contours. Thus when the indeterminism becomes small compared with its initial value, the full adjoint must be used. 

Finally, note that in the PMS local minima can only have $I=0$ \citep[Theorem 2]{2001Judd}, but only one of these minima will correspond to truth. Intuitively, if the direction taken towards a trajectory is the fastest one possible (i.e. by using the true adjoint) then there is a greater probability that the point on $I=0$ it approaches will be closer to the truth than a point reached by moving the state around more in $I$-space. This brings in observational noise as a factor; the smaller the noise the more likely the adjoint is to move the state towards the truth in $I$-space, however the adjoint is approximated.

It is encouraging to note that the two orders of magnitude decrease in indeterminism that \citet[][Fig.~9]{2009Stemler} find for the Lorenz system with an analytical adjoint is comparable with the decrease we find in a system of much greater complexity, even though we use the $\lambda$-adjoint. Before our results move away from the truth around $h=10$, this comparison is also true for the distance between shadow analysis and truth. In our case the distance decreases by a factor of three, as does theirs using an analytical adjoint. When they used the $\lambda$-adjoint they were only able to decrease the distance to truth by a factor of two. Using the $\lambda$-adjoint with a quasigeostrophic model of 1500 variables, \citet{2004JuddB} were able to produce a descended pseudo-orbit that reduced the initial distance from truth by at most a factor of four (their Fig.~3).

\citet{2009Stemler} conclude that the optimal strategy is to use the $\lambda$-adjoint up to a point while noise is removed, before changing to a more accurate approximation to push the sequence towards a trajectory. With an adjoint model for MORALS this could be done, perhaps changing to the full adjoint once the distance from truth reaches a minimum. This strategy would only be possible in the PMS or IMS, of course; it is not obvious how to determine when the switch should occur when the truth is unknown.

\subsection{How the results vary with $\lambda$}
\label{sec:lambda}

For a given model setup and observational noise, $\lambda$ is the primary tunable parameter in the gradient descent. The other parameter, $\Delta\tau$, is optimized by the gradient descent itself. $\lambda$ quantifies the ratio between mismatch information passed backwards and forwards in time. We ran several gradient descents with exactly the same setup, including the same observations, and ten values of $\lambda$: 0, 0.01, 0.04, 0.1, 0.25, 0.35, 0.5, 0.6, 0.8, and 1.

\begin{figure}[tb]
  \centering
	\subfloat[Indeterminism as a function of $\lambda$ at the five descent times listed in the caption.]{\includegraphics[width=0.495\textwidth,clip,viewport=2 8 580 395]{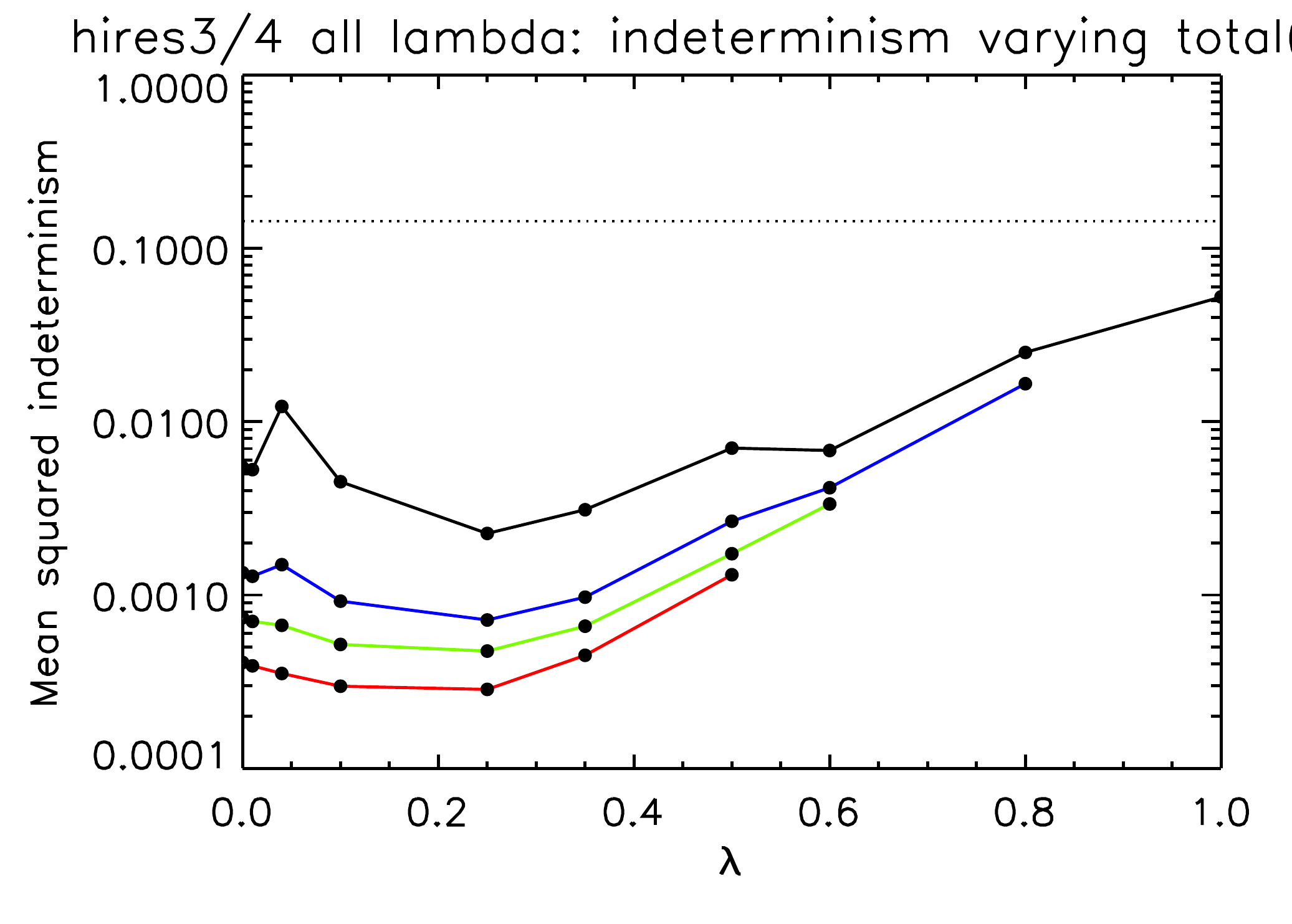}}
	\subfloat[Distance between the sequence of shadow analyses and truth.]{\includegraphics[width=0.495\textwidth,clip,viewport=2 8 580 395]{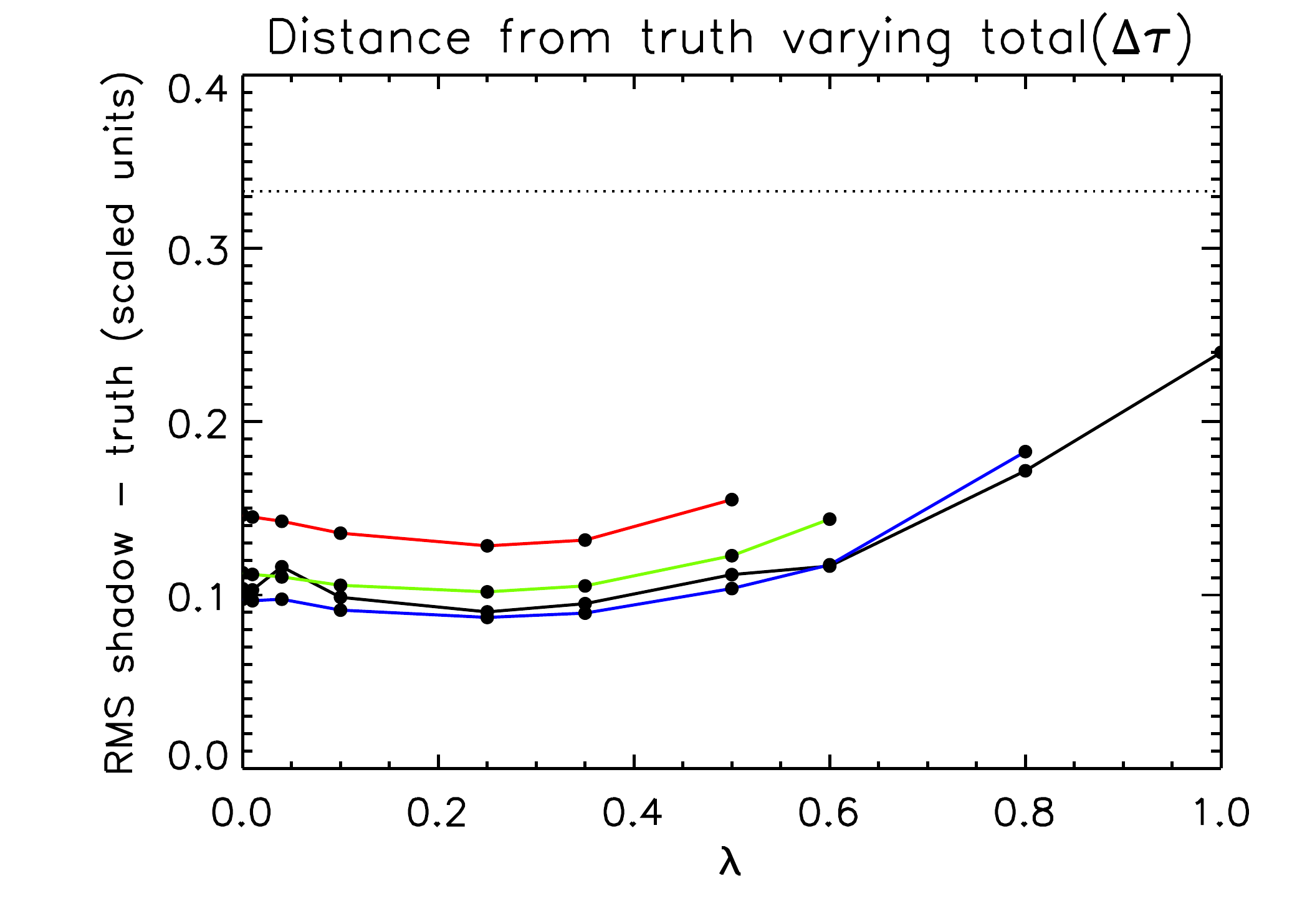}}
	\caption{Progress of the gradient descent (a) indeterminism and (b) distance from truth as a function of $\lambda$. Results are shown at four points during the gradient descent: $\tau=100$ (black), 200 (blue), 400 (green), and 800 (red). The dotted line shows the initial value at $\tau=0$. Large dots show the values obtained; the lines simply join the dots for clarity. The higher values of $\lambda$ are not plotted at some of the later descent times because in these cases the step length $\Delta\tau$, which is optimized by the algorithm, had been halved so many times the descent time did not reach the point plotted even after 500 iterations (in some cases $\Delta\tau$ reached double precision rounding error).}
	\label{fig:indeterminism-lambdas-variation}
\end{figure}

Figure~\ref{fig:indeterminism-lambdas-variation} shows how the indeterminism and distance from truth, the two main diagnostics for the progress of the gradient descent, vary with $\lambda$ during the gradient descent. An intermediate value of $\lambda$ is optimal; both the rate at which indeterminism falls most quickly and the smallest distance to truth occur at intermediate values. There is a smooth variation in both quantities around the minimum at $\lambda=0.25$. Increasing $\lambda$ further degrades both diagnostics: at $\lambda=1$ indeterminism only falls to half its original value by the end, and the distance from truth is substantially larger than for intermediate $\lambda$. In this case the original $\lambda=0.5$ used by other authors is found to be suboptimal, certainly in terms of indeterminism. The distance from truth reaches a minimum before rising again for all $\lambda$. The value of $\tau$ corresponding to the minimum distance during the gradient descent varies only weakly as $\lambda$ is varied. In all cases it occurs around $h\approx10$, between $\tau=100$ and 200. Furthermore, if $\lambda<0.5$ then the depth and position of the minimum are only weakly dependent on $\lambda$. 

\begin{figure}[tb]
  \centering
	\includegraphics[width=0.5\textwidth,clip,viewport=35 15 580 395]{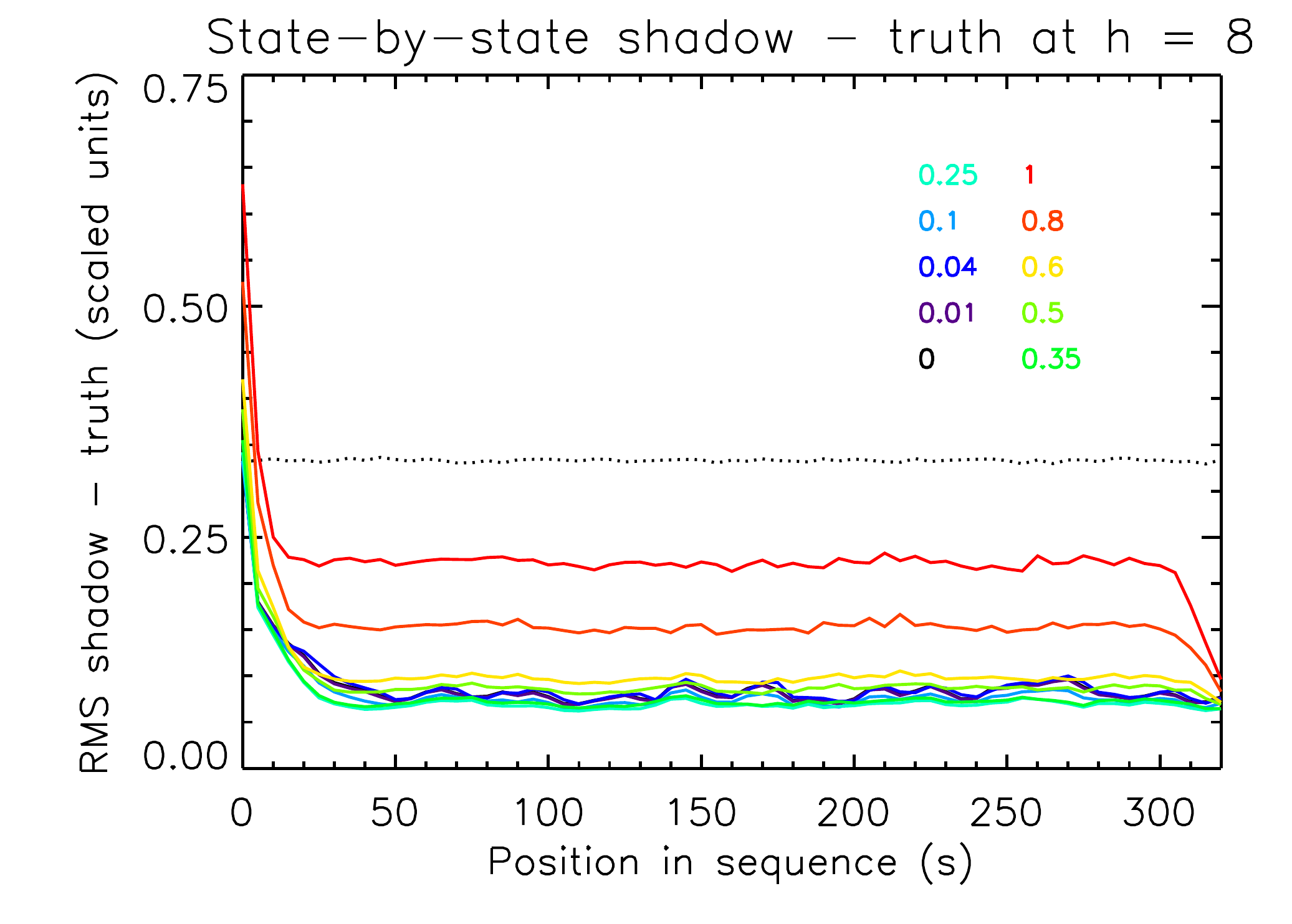}
	\caption{Euclidean distance from each state in the sequence to truth $\Vert({\bf x}_{i,h}-\hat{\bf x}_i)\circ{\rm r}^{-1}\Vert$, given as the mean distance per grid point, after $h=8$ iterations. One line is plotted for each value of $\lambda$; the value for each line is shown in the key. States are separated by 5\,s. The dotted line shows the distance from truth at $h=0$, which is the same for each value of $\lambda$ as it is a function of observations and truth only.}
	\label{fig:truth-dist-all-lambdas-states-8}
\end{figure}

In Fig.~\ref{fig:truth-dist-all-lambdas-states-8} we show how the distance between the shadow analyses and truth varies along the sequence. Except for $\lambda>0.6$, the distance from truth over most of the sequence depends very weakly on $\lambda$. Each line has a similar structure: at the start of the sequence the distance from truth is largest; it falls exponentially to a constant value; remains approximately constant for most of the sequence, and in some cases falls at the very end. The rate of exponential decay at the start of the sequence is fastest for low $\lambda$, and the rate of decrease at the end is fastest for large $\lambda$. From a close inspection it was found that, as above, $\lambda=0.25$ is closest to truth over the part of the sequence where the distance is approximately constant.

The shape of the curves in this plot are very similar to \citet[][Fig.~3]{2004JuddB}. They saw this effect using a quasigeostrophic model and the $\lambda$-adjoint, so this result tells us something more general about the behaviour of the gradient descent using the $\lambda$-adjoint. \citet{2002Ridout} argue that at the start of the sequence the distance from truth will decrease exponentially at a rate given by the closest non-positive Lyapunov exponent to zero, and at the end the distance will increase exponentially at a rate given by the smallest non-negative Lyapunov exponent. At the sequence ends information is only passed in one direction, so there is less information there to guide $\mathcal{X}_h$ towards truth. Neither we nor \citet{2004JuddB} saw the increase at the end of the sequence, however, which they left unexplained. It is not immediately clear why this happens, but perhaps it is because the model itself is not being used to pass information backwards in time. In the $\lambda$-adjoint case the propagation of mismatch backwards in time uses no information about the model's Lyapunov exponents, but this information would be included implicitly in the full adjoint.

Later in the gradient descent than Fig.~\ref{fig:truth-dist-all-lambdas-states-8} the approximately constant distance from truth over most of the sequence gives way to oscillatory functions of position. In fact the distance between shadow analysis and truth after 500 iterations is quite sensitive to the choice of $\lambda$. This change during the later part of the gradient descent implies that during the first iterations the $\lambda$-adjoint removes noise rather than searching for the underlying trajectory. Only later does the gradient descent try to converge towards a trajectory, and the underlying variation along the sequence is revealed.

It is not surprising that $\lambda=0$ does not give the smallest distance from truth. In the trivial case of $\lambda=0$ no information is passed backwards in time so $\mathcal{X}_h$ will eventually converge to a model trajectory starting from the observed state at the start of the sequence. As the system is chaotic this trajectory will diverge exponentially from truth until it is the same order of magnitude as the model attractor width, so we do not expect $\lambda=0$ to approach truth no matter how long the gradient descent is run for. Small but nonzero values of $\lambda$ will also exhibit this effect, but the magnitude of the effect will decrease as $\lambda$ increases. Hence we expect to find the minimum in the distance from truth at nonzero $\lambda$. With indeterminism the effect is similar; at very small $\lambda$ information is predominantly passed along the sequence in one direction, so only the mismatch between states ${\bf x}_{i-1}$ and ${\bf x}_i$ pushes the sequence closer to a trajectory. At larger values the sequence is pushed towards a trajectory from both directions, causing the indeterminism to fall more quickly. Like the distance from truth, this effect will increase as $\lambda$ is increased. 

At the other end of the scale, $\lambda\approx 1$, information is passed in both directions equally. Why, then, do the minima not occur at $\lambda=1$? Here we must consider the effect of approximating the adjoint with a diagonal matrix. If the full adjoint is used the mismatches in both directions contain information about the model at time $t_i$. When the $\lambda$-adjoint is used, however, the information passed backwards in time contains no information about the model at $t_i$, only at $t_{i+1}$ where the mismatch is calculated. Hence the quality of the update at $t_i$ is suboptimal when the $\lambda$-adjoint is used. Hence we might expect the quality of the update to increase as $\lambda$ is reduced and more weight in the update step is assigned to the model at $t_i$. Our results show empirically where the balance is between these two effects, at least in the annulus context, and we expect such a trade-off to exist for other chaotic systems, for the same reasons.

The conclusions from this section are clear. First, there is a range of intermediate $\lambda$ values which give reasonable results both in terms of indeterminism and distance from truth, while at both extremes of the range the quality of the gradient descent is compromised. Second, the distance from truth as a function of position in the sequence confirms something more general about using the $\lambda$-adjoint, given the comparison of our results with \citet{2004JuddB}. There is a trade-off between two mechanisms that degrade the quality of the gradient descent for extreme values of $\lambda$. For $\lambda>0.5$ the quality of the update step is degraded by the relatively large weight assigned to information passed backwards in time sub-optimally. $\Delta\tau$ also falls very quickly with high $\lambda$, so only a small amount of descent time is covered, limiting the potential of the gradient descent to proceed much further. For $\lambda<0.1$ the sequence converges to a trajectory starting from very close to the first observation, so chaos causes the rest of the shadow analyses to diverge from truth. For intermediate values of $\lambda$, where the combined effect is minimized, sequences are recovered closest to a trajectory and to truth. We recommend a value around 0.25 for future applications using the $\lambda$-adjoint, although within this intermediate range the results are only weakly dependent on $\lambda$.

\begin{figure}[tbp]
  \centering
  \includegraphics[width=0.82\textwidth,clip,viewport=10 100 820 510]{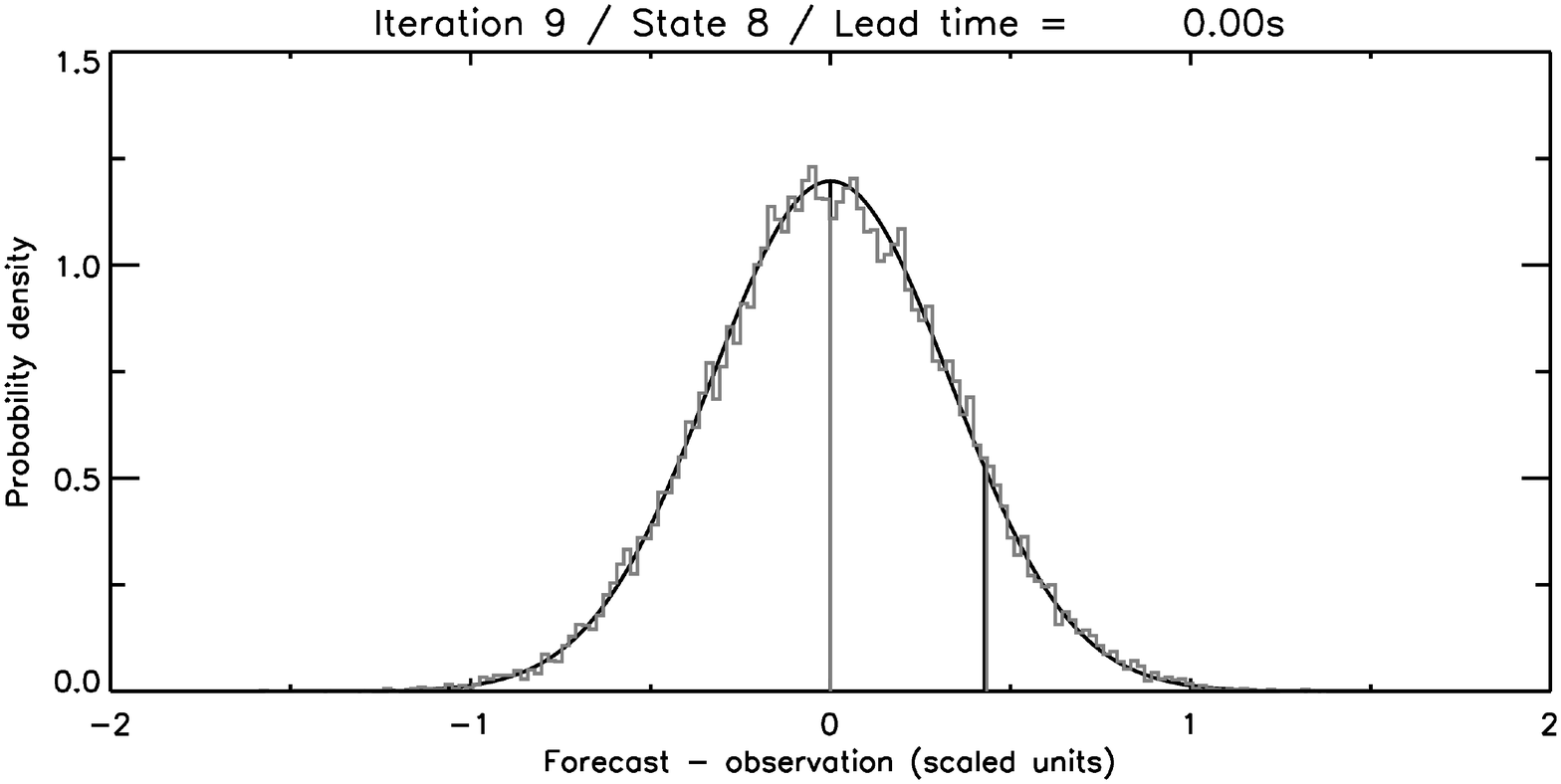}\\$ $\\
  \includegraphics[width=0.82\textwidth,clip,viewport=10 100 820 510]{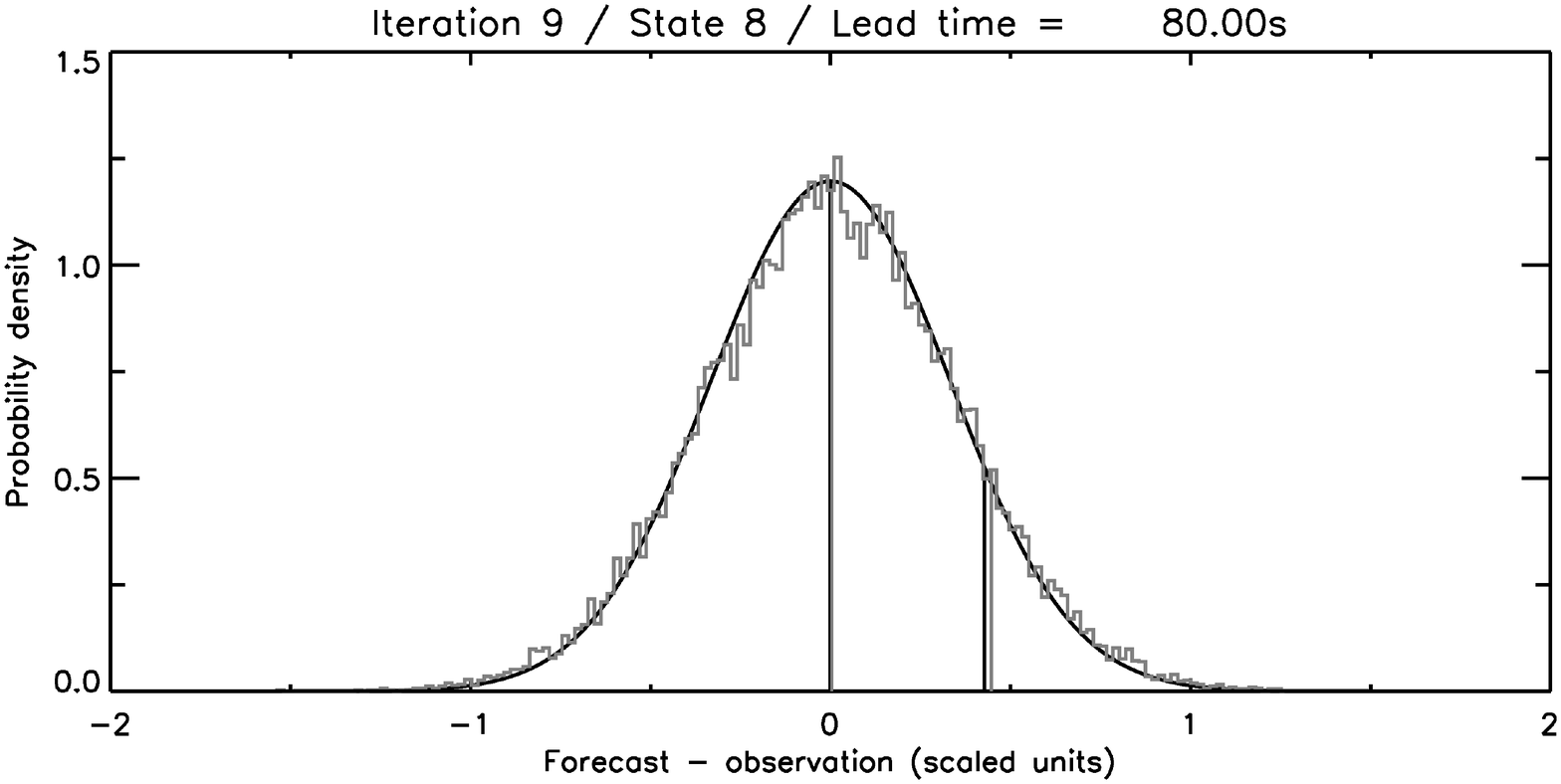}\\$ $\\
  \includegraphics[width=0.82\textwidth,clip,viewport=10 100 820 510]{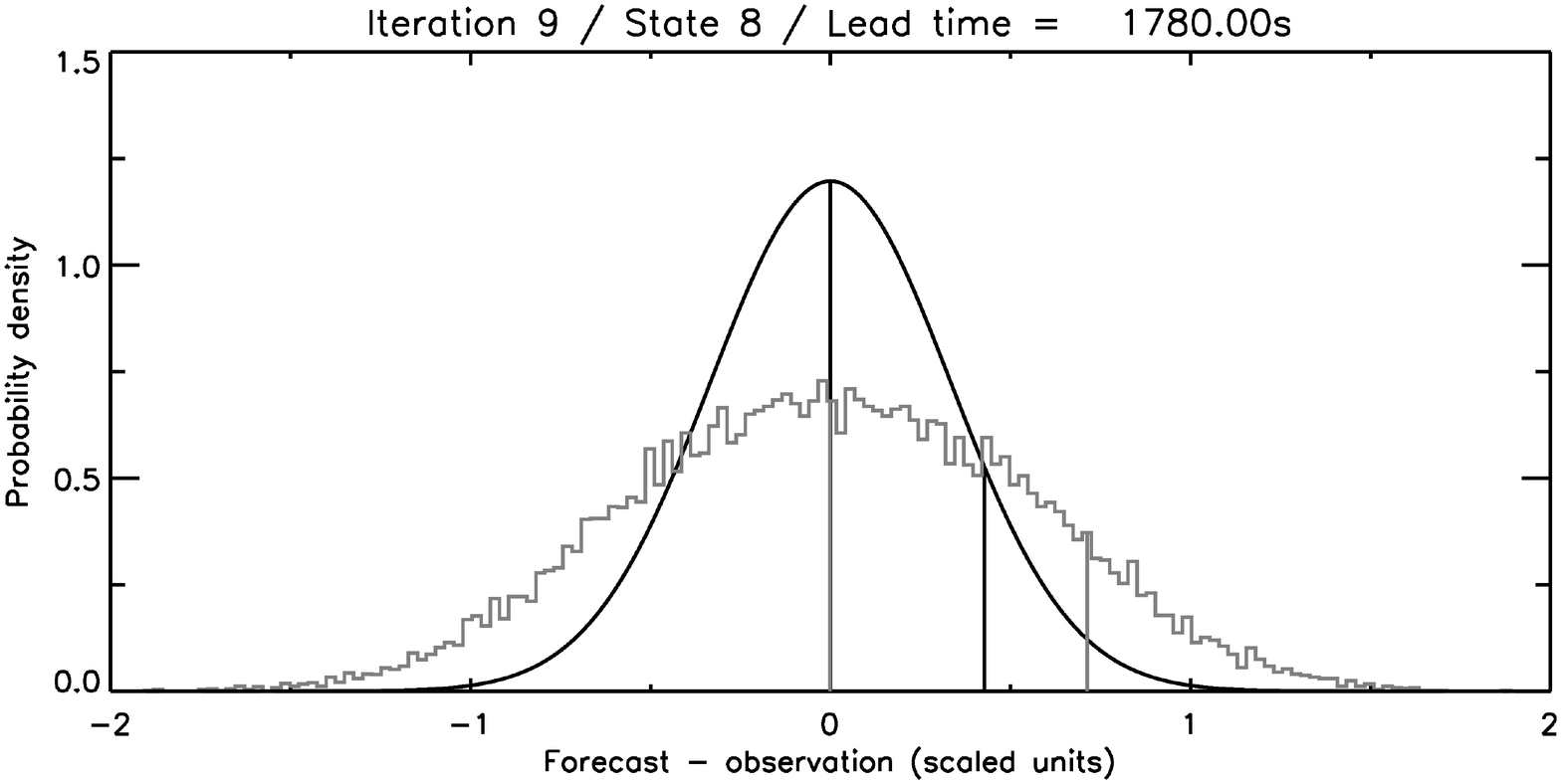}
  \caption{Distributions of residual errors ${\bf e}[t]$ (Eq.~\ref{eq:forecast-errors}) at three lead times from the candidate trajectory started from state $i=8$ at $h=9$ for the gradient descent with $\lambda=0.25$. Top: $t=0$, middle: $t=80$\,s (the first time the trajectory fails to shadow observations at significance level $p=10^{-5}$), and bottom: $t=1780$\,s (the end of the candidate trajectory). The black and grey curves are the noise and residual distributions, and the vertical black and grey lines are the 50th and 90th percentiles of the noise and residual distributions. Notice how strict this definition is --- even in the middle panel the residual distribution is only marginally different from the noise distribution, yet the model does not shadow at this time.}
  \label{fig:hires4a-distributions}
\end{figure}

\section{Shadowing times from the sequence of shadow analyses}
\label{sec:shadowing}

Gradient descent produces, from a sequence of observations, a pseudo-orbit of the model closer to a trajectory than the original observation sequence. Each state on the pseudo-orbit is the start of a candidate trajectory of the model. The shadowing time is the maximum time any of these candidates shadow the observations. We can also generate additional candidates using linear combinations of states on the pseudo-orbit and forecast images of earlier states. 

In this section we measure how long the model can shadow observations using candidates from a pseudo-orbit produced by gradient descent in the previous section, each candidate beginning a model trajectory. The \textit{shadowing time} $\tau_{S_t}$ is the maximum among all \textit{candidate shadowing times} $\tau_S$ starting from candidates $\bf x$, i.e. $\tau_{S_t}=\max_{\bf x}\tau_S({\bf x},t)$. From the previous section we choose candidates from the gradient descent with $\lambda=0.25$ at iteration $h=9$, because at that iteration that gradient descent came closest to truth among all the cases that were run. \citetalias{2019YoungQ} describes the method used to compute the shadowing time; we summarise it below.

Our shadowing time quantifies how well the gradient descent performs against a benchmark set in \citetalias{2019YoungQ}. In that work they used the same model parameters as in this paper, except here we have used $\sigma=1/3$ instead of $\sigma=0.1$. They generated a cloud of candidate initial conditions a fixed distance from the true state and for each candidate measured how long it shadowed a subsequent set of observations. These times provide a benchmark against which we can compare the shadowing times from our gradient descent experiments --- the shadowing times we might expect depend on the distance of the candidates from the original true trajectory. The largest initial distance from truth that \citetalias{2019YoungQ} used was about 17.5\% of the observational error, and in that case they measured shadowing times of 50--150\,s. In the previous section we found the distance from truth of our best-case result is around 25\% of the observational error (Fig.~\ref{fig:indeterminism-lambdas-variation}), so we might expect a maximum shadowing time around 100\,s.

From the selected pseudo-orbit we took as candidates at each $t_i$ (1) the state on the pseudo-orbit ${\bf x}_{i,h}$, (2) the state halfway between this state and the forecast image of the previous state, $\frac{1}{2}[{\bf x}_{i,h}+f({\bf x}_{i-1,h})]$, and (3) the images of all the previous states on the pseudo-orbit mapped to $t_i$, for both ${\bf x}_{i,h}$ and the halfway state. The shadowing times for the candidates (3) are available at no extra computational cost. Each candidate was used to start a single MORALS trajectory. 

\subsection{Measuring the shadowing time}

Following \citet{2010Smith}, our candidate shadowing time $\tau_S$ for a particular candidate trajectory is the trajectory length over which the residual error distribution remains consistent with the observational error distribution. The vector of residual errors is
\begin{linenomath*}\begin{equation}
  {\bf e}[t]=\left\{f^t({\bf x}[0])-{\bf s}[t]\right\}\circ{\bf r}^{-1}
  \label{eq:forecast-errors}
\end{equation}\end{linenomath*}
where ${\bf x}[0]$ is the initial candidate state, ${\bf s}[t]$ are the observations at time $t$ from the beginning of the candidate trajectory, and $f^t$ denotes integration of the model for time $t$. We scale the raw residual error by $\bf r$ so we can combine different physical quantities into one distribution. We test the null hypothesis that the vector ${\bf e}[t]$ is a sample drawn from the observational error (noise) distribution $N(0,\sigma^2)$ (the comparable distribution once raw values are scaled by $\bf r$). \citetalias{2019YoungQ} showed that, using order statistics, the distribution of a specific percentile of the noise distribution can be found analytically. We find (empirically) the 50th and 90th percentiles of the residual distribution ${\bf e}[t]$, and test whether, at a particular significance level $p$, these are drawn from the equivalent noise distribution (found analytically). For the candidate trajectory to shadow the observations at time $t$ we require \textit{both} percentiles to fall within the respective confidence intervals of the noise distribution. 

If the model shadows the observations at time $t$ by this definition, we proceed to the next observations (5\,s later in this case) and repeat the procedure. The shadowing time $\tau_S$ for a particular candidate is the last time at which the residual error distribution is consistent with the noise distribution. Figure~\ref{fig:hires4a-distributions} shows this definition in use.

We require a suitable significance level $p$ to find $\tau_S$. Because our shadowing time algorithm requires multiple significance tests, the probability of a Type I error (rejecting the null hypothesis when it is true, and hence setting an erroneously low $\tau_S$) increases as more tests are done. \citetalias{2019YoungQ} calculated the maximum $p$ leading to fewer than one expected Type I error over all candidates. The calculation is outlined in \ref{sec:significance}; we found $p=10^{-5}$ to be sufficient.

\subsection{Measured shadowing times}

\begin{figure}[tbp]
  \centering
  \includegraphics[width=0.7\textwidth,clip,viewport=60 20 840 393]{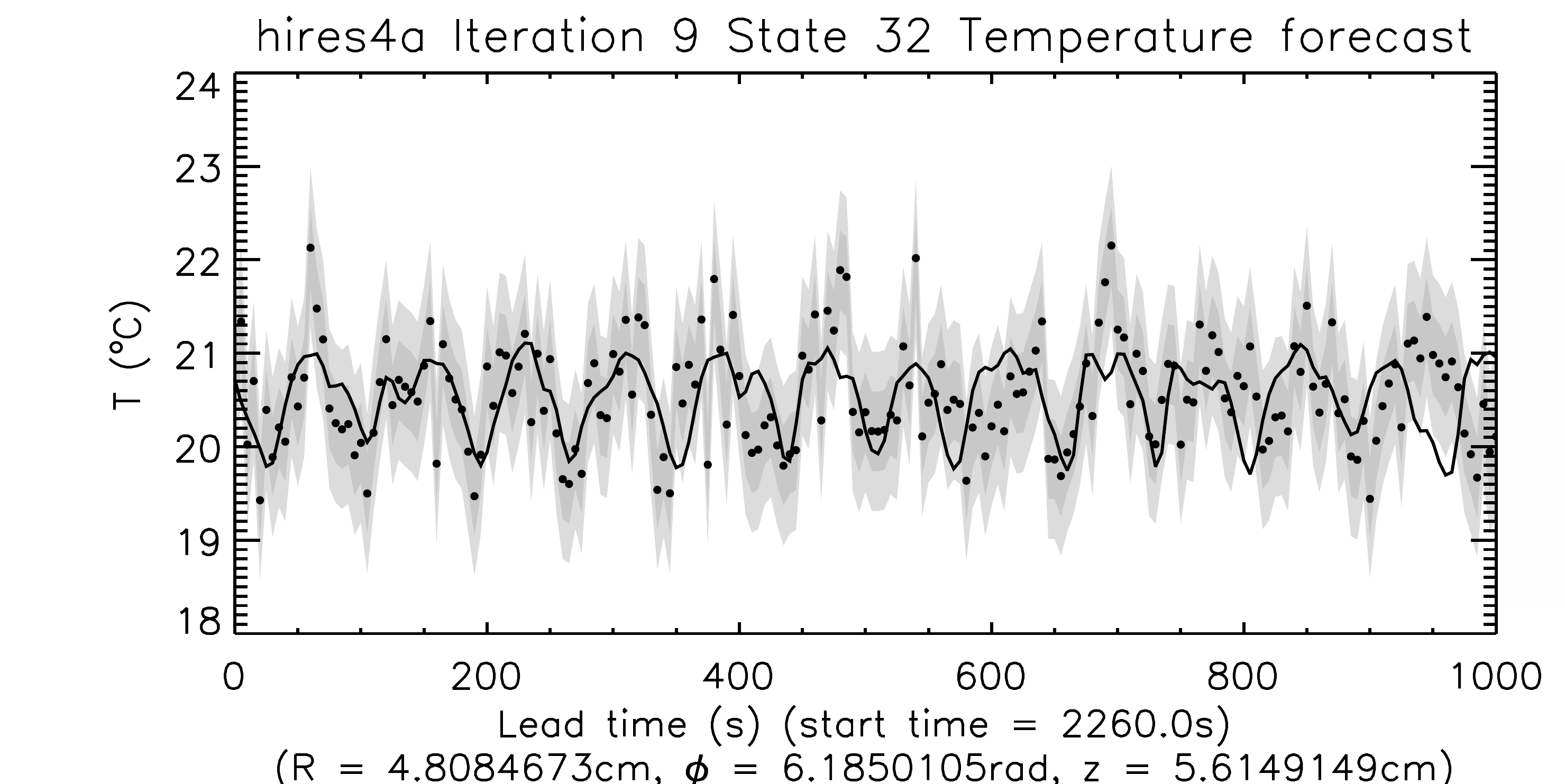}
  \caption{Temperature time series at the single grid point used earlier in the paper (Fig.~\ref{fig:descent-single}) for the candidate trajectory started from state $i=32$ in the sequence at $h=9$ with $\lambda=0.25$. The solid black line shows the forecast, black dots show the observations, and the shaded area represents observations $\pm 1\sigma$ (darker shading) and $\pm 2\sigma$ (lighter shading).}
  \label{fig:hires4a-tseries}
\end{figure}

\begin{figure}[tb]
  \centering
  \includegraphics[width=0.6\textwidth,clip,viewport=40 20 580 391]{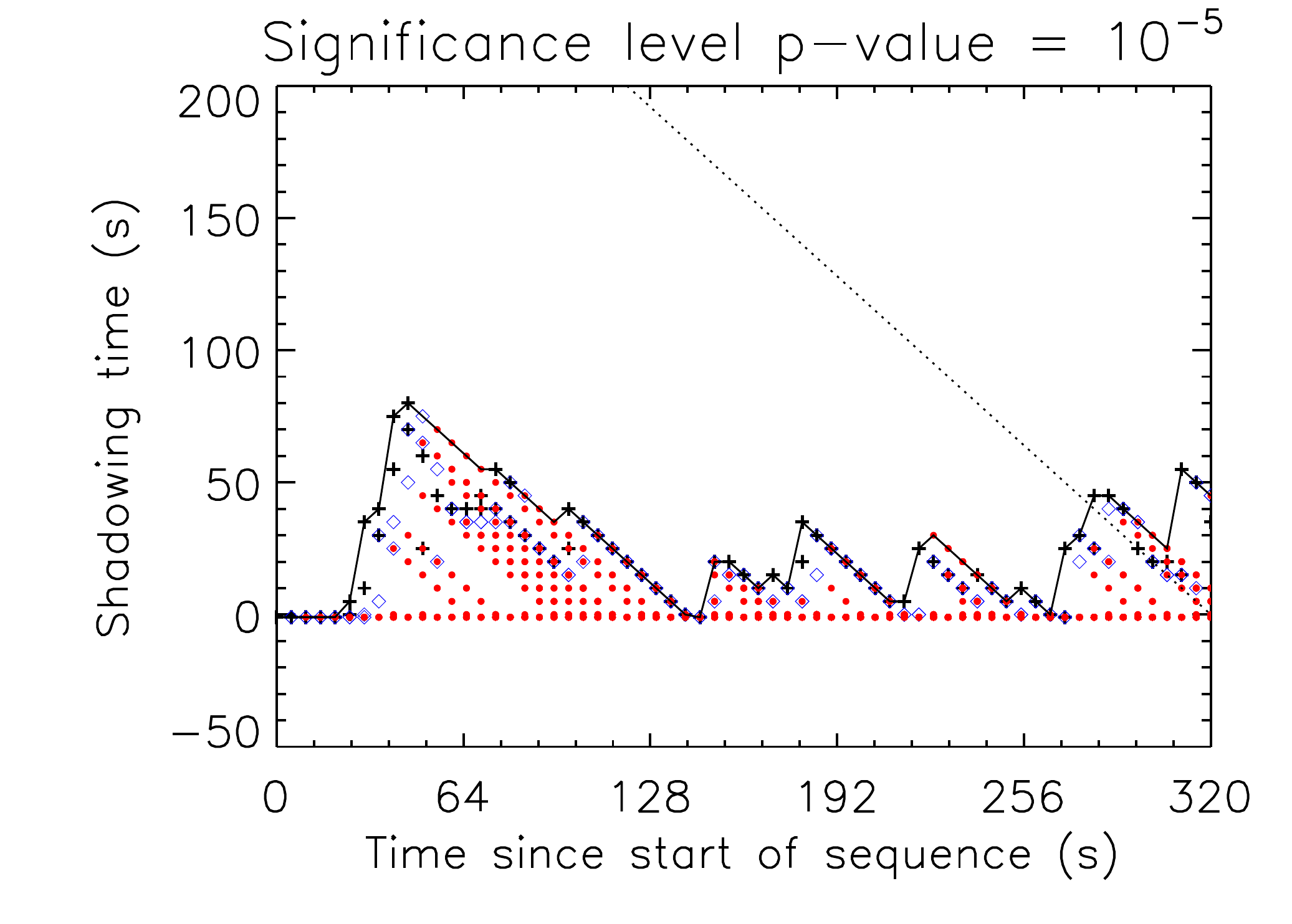}
  \caption{Shadowing times $\tau_S$ for each of the candidates described in the text, for the pseudo-orbit after $h=9$ in the gradient descent with $\lambda=0.25$. At each position in the sequence, crosses (black) are candidates ${\bf x}_{i,h}$ and $\frac{1}{2}[{\bf x}_{i,h}+f({\bf x}_{i-1,h})]$, diamonds (blue) are candidates from one-step forecast images of the previous step, i.e. $f({\bf x}_{i-1,h})$ and $f\left(\frac{1}{2}[{\bf x}_{i-1,h}+f({\bf x}_{i-2,h})]\right)$, and dots (red) are candidates from forecast images of all previous states. The solid line traces the maximum $\tau_S$ over all candidates at each point in the sequence, and the dotted line separates candidates that shadow to the end of the original sequence of observations (above the line) from those that don't (below it).}
  \label{fig:hires4a-consistency}
\end{figure}

We measured shadowing times $\tau_S$ for each of the candidates. Figure~\ref{fig:hires4a-tseries} shows an example time series from one of the candidate trajectories, and Fig.~\ref{fig:hires4a-consistency} shows all the candidate shadowing times $\tau_S$. From this plot we can read off $\tau_{S_t}=80$\,s as the shadowing time for this pseudo-orbit of descended states. This is about one quarter of the total length of the original observational sequence, or about one period of the longest timescale of the system, the oscillation of the main baroclinic wave. This time is near the low end of the range of shadowing times obtained in \citetalias{2019YoungQ} for their largest perturbation from truth, but the descended pseudo-orbit's distance from truth in this case was slightly larger, around 25\% of the observational error compared with 17.5\%.

There are no particular trends relating individual candidates' shadowing times to their positions in the sequence, except the candidates starting from very close to the start of the sequence do not shadow at all. This is not surprising since the quality of the states near the start of the sequence are poorer than in the middle because information is only passed in one direction at the start of the sequence. A few observational states appear to be difficult for the candidate trajectories to shadow (e.g. around 140--145\,s), causing all candidate trajectories approaching it to fail to shadow at that point. Within 45\,s of the end of the sequence we found candidates that shadow to the end of the sequence.

Our gradient descent setup and algorithm is really just a method for selecting candidate trajectories we think will shadow for a long time. It is always possible that we have missed candidates that shadow for longer, so our measured shadowing time $\tau_{S_t}$ can only be a lower bound. Indeed, we know (by construction) that at least one candidate exists that shadows the whole observational sequence: the true sequence that generated the observations in the first place. Clearly our $\tau_{S_t}$ does not approach this, being only about a quarter of the original sequence length. However, our shadowing time is consistent with those for candidate trajectories starting from a similar distance from the truth in \citetalias{2019YoungQ}. In this respect our results are encouraging, because they show that the distance from the candidate state to truth is the primary predictor of the subsequent candidate shadowing time. It shows that better shadowing times will come from improvements to the gradient descent method that bring the pseudo-orbit closer to truth. We identified some possible improvements in the previous section, foremost among these being to use a full adjoint model instead of the na\"{i}ve $\lambda$-adjoint. With a full adjoint we expect to shadow for considerably longer.

\section{Discussion and conclusions}
\label{sec:conc}

We have implemented gradient descent of indeterminism for the thermally-driven rotating annulus in the perfect model scenario. Our results show that a sequence of states much closer to a true system trajectory can be recovered using gradient descent. Diagnostics based on indeterminism and distance from the truth showed our demonstration gradient descent recovered a sequence of states in which the indeterminism had fallen by two orders of magnitude. The sequence converged towards truth as the gradient descent progressed but then moved away from truth once the distance had fallen by a factor of three. This was attributed to the $\lambda$-adjoint approximation. An analysis of varying $\lambda$ showed that the gradient descent is optimized around $\lambda=0.25$. In that case indeterminism falls by three orders of magnitude after 500 gradient descent steps and the distance from truth falls by a factor of four except near the start of the sequence.

Candidate trajectories started from one particular $\mathcal{X}_h$ were used to obtain shadowing times using the method developed in \citetalias{2019YoungQ}. We found the model shadows the observations for $\tau_{S_t}=80$ at the $p=10^{-5}$ significance level. This was at the lower end of the range of times obtained in \citetalias{2019YoungQ} for candidate trajectories started a similar distance from truth. Our shadowing time is encouraging because the initial distance between $\mathcal{X}_h$ and truth could be decreased further by using a more accurate adjoint model for the gradient descent.

We discussed above how the $\lambda$-adjoint causes the sequence of shadow analyses to move away from truth after the first several steps of the gradient descent. Despite this result, we are most encouraged that, even with the $\lambda$-adjoint, the distance from truth can be reduced by a factor of four over most of the sequence. The most important next step in this work is to include a full adjoint model for MORALS. At the time the algorithm was implemented no adjoint model was available, but one now exists \citep{2010Hussain}, which should be a major step forward in using this model for various purposes. This improvement may be of greatest benefit in the PMS, as in experiments with observational data \citet{2008JuddA} demonstrated that the $\lambda$-adjoint may be sufficient because the advantages of a full adjoint in the later part of the gradient descent are offset by the model being an imperfect representation of the system. The attractors of the model and system will be disjoint, and so there may be little to gain from a full adjoint when its main advantage is to navigate through the complex structure of the indeterminism contours near the model attractor. Nevertheless, an accurate adjoint with an imperfect model beats an inaccurate adjoint with an imperfect model. Any increased accuracy gained with the full adjoint using real observations should be measured and compared with the $\lambda$-adjoint. Quantification of this increased accuracy would be useful for informing operational implementation of any algorithm based on gradient descent methods. If both the $\lambda$-adjoint and full adjoint are used, it is not clear when the switch over should occur when the true state is not available. One option is to switch over when the distance between $\mathcal{X}_h$ and $\mathcal{X}_0$ (the \textit{implied noise}) first reaches a maximum.

We are excited by the possibility of applying this method to real annulus laboratory data. Gradient descent allows model error to be examined in a systematic way, by examining the geometric relationship between observations and the model attractor \citep{2008JuddA}. Application of this method to laboratory data will allow model error to be explored in more detail, leading to a more fundamental understanding of the limitations of our annulus model. When using observational data the visual progression of the gradient descent (i.e. Fig.~\ref{fig:descent-slices}) is even more revealing, as then one sees how the model adjusts itself where the original sequence of observations or analyses is far from the model manifold. This reveals locations and features in the flow that are poorly simulated by the model, and also how the model attempts to adjust itself to fit to those observations. One particular annulus dataset that might be of particular interest is a wavenumber-3 structural vacillation flow where the observed wave is phase-locked to the tank because of the deposition of tracer particles onto the bottom. This flow is generally poorly modelled by MORALS \citep{2013Young}, and using this dataset with gradient descent should provide some insight into how the model fails in this case.

The broader aim is to determine whether gradient descent would be feasible in an operational context. For this to be the case, it would (at least) need to out-perform the current state-of-the-art data assimilation technique used in forecasting centres worldwide, 4D-Var \citep{2007Rawlins}, and out-perform the sequential Bayesian methods in development such as the ensemble Kalman filter (EnKF) \citep{1994Evensen} and particle filters \citep{2010vanLeeuwen}. While a numerical comparison is beyond the scope of this paper, there are a number of conceptual advantages of gradient descent over these other methods. These differences are discussed in more detail by \citet{2009Stemler} and \citet{2010Judd}. 

Sequential Bayesian methods like the Kalman filter and its variants suffer from the major problem that the observational noise must be small compared with the system's nonlinearities. Sequential methods also cannot correct poor state estimates in the past whose error then propagates forward \citep{2010Judd}. For nonlinear systems in particular the forward propagation of errors can introduce large errors very quickly. Gradient descent avoids these problems with sequential filtering as it makes no assumptions about the linearity of the system. Indeed, nonlinearity is actively exploited. In addition, it uses information from the past simultaneously with information from the present in each state estimate. \citet{2003Judd} showed that gradient descent compares favourably with the extended Kalman filter (EKF), except when the dynamical noise exceeds observational noise. An additional problem with these and any methods based on maximum likelihood estimates is that even in the PMS these methods fail to assign that maximum likelihood to the true state of the system \citep{2007JuddA}. Comparing gradient descent with particle filtering using the \citet{1979Ikeda} system, \citet{2009Judd} found that gradient descent recovers state estimates closer to truth than the particle filter in almost all cases, even when using an optimized particle filter and ``out-of-the-box'' gradient descent.

Of the variational methods 4D-Var is, in a sense, also a method that searches for shadowing trajectories. The fundamental difference is that 4D-Var uses a kind of ``shooting'' method; it alters the initial state in a sequence and compares the model trajectory generated from that initial state over a window of observations. Instead of indeterminism the cost function is \citep[Eq.~8]{2009Stemler}
\begin{linenomath*}\begin{equation}
  C(x)=\frac{1}{n}\dsum_{i=1}^{n}\Vert s_i-f^i(x)\Vert^2
\end{equation}\end{linenomath*}
The problem with this method is that sensitivity to initial conditions means that the window length over which 4D-Var can be realistically applied is severely restricted. Gradient descent suffers from no such problem as states all the way along the window are used simultaneously, and hence each forecast needs to be optimized only over the time between it and the next state in the sequence. \citet[][Fig.~8]{2009Stemler} demonstrate this problem with 4D-Var using the \citet{1963Lorenz} system. A related variational method, weakly constrained 4D variational assimilation (WC4DVA), also has some similarities to gradient descent. \citet[p.~1268]{2009Stemler}, \citet[p.~221]{2008Judd}, and \citet[pp.~268--9]{2010Judd} argue strongly, however, that these similarities are superficial. In particular, they show that there is an inconsistency between what WC4DVA claims to solve and what the method actually solves.

Gradient descent offers a number of additional practical advantages over these other methods. First, the algorithm is, in our opinion, conceptually simpler than variational or EnKF methods. Second, when using real data the number of tunable parameters is generally less than other assimilation methods. Analysis correction \citep{1991Lorenc} has about ten, for example. With a full adjoint the only tunable parameter in gradient descent is $\Delta\tau$, and even then its value can be optimized by the gradient descent as described in \ref{sec:gd-morals}. Third, the background error covariance matrix is not required, the calculation of which is generally a major challenge for other methods. Model errors are ``discovered, not prescribed'' \citep{2008JuddA,2010Judd}, providing information about where and how the model fails to simulate reality.

The main problem we found using gradient descent in this system was the burden on computational resources. Each gradient descent step requires the resources for one complete pass through $\mathcal{X}_h$ with both the forward model and the adjoint model. In the PMS this is a major problem as many hundreds of iterations are required. This would not be such a problem with laboratory data, however, because models of real systems are imperfect. The gradient descent is therefore not expected to converge to a trajectory of the model. Instead it will converge to a pseudo-orbit and in practice the gradient descent is terminated when the standard deviation of the forecast mismatches approaches observational error. This only takes some tens of steps, which is a great improvement over the hundreds of steps required in the PMS. Even so, tens of iterations of a GCM over a lengthy sequence of observations is somewhat more computation than is currently used in operational assimilation. For example, the Met Office 4D-Var scheme uses one pass of the nonlinear model and 5--6 passes of the linearized model through the sequence \citep{2007Rawlins}. For gradient descent to be a feasible operational method, therefore, it would need to be shown that its additional accuracy is worth the computational expense.

With these comparisons in mind, we feel there is great potential for using gradient descent for state estimation in high-dimensional models, and in particular the rotating annulus' part in developing and testing the method. The framework for using gradient descent in the annulus context is now in place. In the future one could extend it to estimates of shadowing times in the various annulus flow regimes, experiments in the imperfect model scenario, and experiments starting the gradient descent from laboratory data. Several questions present themselves: How long can the laboratory annulus be shadowed in different flow regimes? How long a window is required to shadow with the same accuracy as sequential methods (in particular analysis correction; \citet{2013Young} have produced a set of rotating annulus assimilation results for comparison)? How quickly does the gradient descent converge, and is it quicker when an analysis correction analysis is used to begin the gradient descent? How do these results depend on model resolution, and  is there a resolution beyond which no further improvements can be made? 

Finally, we intend to use gradient descent as part of a larger programme of research using the annulus as a test bed for meteorological methods in current use and development. As mentioned above, analysis correction has already been implemented by \citet{2013Young} as an example of a well-established assimilation method. We intend to implement and compare some of the other methods discussed in this work in the annulus context, for example 4D-Var and the particle filter; \citet{2010Ravela} have already made some progress with the EnKF. Shadowing methods also facilitate ensemble generation using the theory of indistinguishable states \citep{2001Judd,2004JuddA}, and we believe the application of this particular method in a real physical system would also be a timely comparison to make with current methods for ensemble generation.

\section*{Acknowledgments}

We thank Daniel Bruynooghe, Hailang Du, Kevin Judd, and Thomas Stemler for useful conversations on a number of topics. RMBY and FN acknowledge financial support from the Grantham Research Institute on Climate Change and the Environment. RMBY also acknowledges financial support from NERC Studentship NER/S/A/2005/13667.

\appendix

\section{Technical details}

\begin{table}[tb]
\centering
  \caption{Annulus and MORALS parameters. Fluid properties are parameterized as a function of temperature using the expressions in \citet[Table 1]{1985HignettB}.}
  \begin{tabular}{lll}
	 \toprule
	 Inner cylinder radius & $a$ & 2.5\,cm\\
	 Outer cylinder radius & $b$ & 8.0\,cm\\
	 Annulus depth & $d$ & 14.0\,cm\\
	 Rotation rate & $\Omega$ & 1.00\,rad\,s$^{-1}$\\
	 Reference temperature & $T_{\rm R}$ & 22\,$^{\circ}$C\\
	 Inner cylinder temperature & $T_a$ & 18\,$^{\circ}$C\\
	 Outer cylinder temperature & $T_b$ & 22\,$^{\circ}$C\\
	 Temperature difference & $\Delta T$ & 4\,degC\\
	 \midrule
	 Fluid & & 17\% glycerol / 83\% water by volume\\
	 Density & $\rho_0$ & 1.043\,g\,cm$^{-3}$ at 22\,$^{\circ}$C\\
	 Viscosity & $\nu_0$ & 0.0162\,cm$^2$\,s$^{-1}$ at 22\,$^{\circ}$C\\
	 Thermal diffusivity & $\kappa_0$ & 0.00129\,cm$^2$\,s$^{-1}$ at 22\,$^{\circ}$C\\
	 \midrule
	 Model timestep & $\delta t$ & 0.02\,s\\
	 Radial grid points & $N_R$ & 16\\
	 Azimuthal grid points & $N_{\phi}$ & 32\\
	 Vertical grid points & $N_z$ & 16\\
	 \bottomrule
  \end{tabular}
\label{tab:params}
\end{table}

\subsection{MORALS}

MORALS solves the Navier-Stokes, heat transfer, and continuity equations subject to the Boussinesq approximation, in cylindrical polar coordinates. Four prognostic variables are defined: three velocity directions $\bf u$ (radial), $\bf v$ (azimuthal), $\bf w$ (vertical), and temperature $\bf T$. A fifth field required in the prognostic equations is kinetic pressure ${\bm\Pi}\equiv {\bf p}/\rho_0$, which is diagnostic and is calculated from the other four fields using a Poisson equation. $\rho_0$ is the fluid density at a reference temperature. The fluid rotates at constant angular velocity $\Omega$, all velocities are set to zero at the boundaries, the temperature gradient is zero across the top and bottom boundaries, and the temperatures at $R=a$ and $R=b$ are $T_a$ and $T_b$ respectively. $\bf T$ is defined relative to a reference temperature $T_R$ (22\,$^{\circ}$C here) and $\bm\Pi$ is relative to a reference pressure $\Pi_0(R,z)=\frac{1}{2}\Omega R^2+g(d-z)$. The fields are discretized on a staggered Arakawa C grid \citep{1977Arakawa}, and are non-uniform in the radial and vertical directions to resolve the boundary layers.

With four prognostic variables using the resolution in the table there are $N_{\rm tot}=4N_RN_{\theta}N_z=32768$ variables in total. The number of independent variables, $N$, is less than this because points on and outside the fluid boundary (outside points are required by some boundary conditions) are fixed by values at other grid points. We use $N$ as the dimension of the model, $N=24192$. When working in the PMS and correspondence with the laboratory experiment is less important, the choice of resolution depends on a balance between the available computer resources and the number of simulations required for that experiment. Because of the large computational overhead required by gradient descent, we have used a lower resolution than is normally used for annulus simulations that are compared with laboratory observations.

\subsection{Experimental parameters}

Table~\ref{tab:params} lists the annulus and MORALS parameters. Table~\ref{tab:expt} lists the experimental parameters for the demonstration gradient descent in Sect.~\ref{sec:results}.

\begin{table*}[tb]
  \centering
  \caption{Experimental parameters for the demonstration gradient descent in Sect.~\ref{sec:results}.}
  \begin{tabular}{lll}
	 \toprule
	 \multicolumn{3}{c}{Demonstration gradient descent - truth and observations}\\
	 \midrule
	 Spin-up time & $t_{\rm spinup}$ & 2000\,s\\
	 Pre-sequence time & $t_{\rm preseq}$ & 100\,s\\
	 Time between states & $\Delta t$ & 5\,s\\
	 Window width & $w$ & 64 \\
	 Time of first state ($i=0$) & $t_0$ & 2100\,s\\
	 Time of final state ($i=64$) & $t_w$ & 2420\,s\\
	 Sequence length & $t_w-t_0$ & 320\,s\\
	 Observational noise & $\sigma$ & $1/3$\\
	 \midrule
	 \multicolumn{3}{c}{Demonstration gradient descent - gradient descent parameters}\\
	 \midrule
	 Initial step length & $\Delta\tau(\tau=0)$ & 16.0\\
	 Cut-off indeterminism & $\epsilon$ & $10^{-28}$ (machine precision)\\
	 Gradient-free descent parameter & $\lambda$ & 0.5 \\
	 Maximum number of iterations & $h_{\rm max}$ & 500\\
	 \bottomrule
  \end{tabular}
  \label{tab:expt}
\end{table*}

\subsection{Generating the initial sequence of observations}
\label{sec:obs-seq}

A full MORALS state $\bf x$ is a concatenation of the four fields $u$, $v$, $w$, and $T$:
\begin{linenomath*}\begin{equation}
   {\bf x}\equiv\left(\begin{array}{l}
   {\bf u}\\
   {\bf v}\\
   {\bf w}\\
   {\bf T}
   \end{array}\right)
   \label{eq:four-fields}
\end{equation}\end{linenomath*}
where $\dim({\bf x})=N=N_u+N_v+N_w+N_T$. 

The sequence of artifical observations $\mathcal{X}_0$ was generated by adding noise $\tilde{\bf e}_i$ to the true states $\hat{\bf x}_i$ generated by MORALS. The method is very similar to the method used in \citetalias{2019YoungQ}. The true sequence is generated by MORALS in three stages: (1) from $t=0$ to $t_{\rm spinup}$, (2) from $t=t_{\rm spinup}$ to $t_0=t_{\rm spinup}+t_{\rm preseq}$, and (3) from $t=t_0$ to $t_w$. The third stage corresponds to $\hat{\mathcal{X}}$.

The first stage (the ``spin-up'' phase) is required to spin up the model from rest (in the rotating frame of reference) to a state in which transient behaviour has decayed and a coherent flow structure is present.

The third stage (the ``sequence'' phase) contains the sequence of states $\hat{\mathcal{X}}$ used to obtain observations to start the gradient descent. The states are separated in time by $\Delta t$. 

The second stage (the ``pre-sequence'' phase) is used purely for the generation of observations in the PMS. We use the sequence of states between $t=t_{\rm spinup}$ and $t=t_w$ to obtain an estimate, at each model grid point, of the range of values admitted by the model when it is in dynamical equilibrium. This range can be interpreted as a measure of the natural variability of the system at each point in space. We denote this range by the vector ${\bf r}$, where each element represents the range of values that encloses 99\% of the values over the pre-sequence and sequence phases at that grid point:        
\begin{linenomath*}\begin{equation}
  {\bf r}={\bf \tilde{x}}[t_{\rm spinup},...,t_w]_{99.5\%}-{\bf \tilde{x}}[t_{\rm spinup},...,t_w]_{0.5\%}
  \label{eq:r}
\end{equation}\end{linenomath*}

We use an additional period of time before the sequence phase to ensure that the sequence of states used to calculate the scaling is long enough to avoid problems of small-number statistics, while remaining short enough to avoid including variability on longer time scales than are represented in the sequence. In practice the pre-sequence was about 30\% of the length of the sequence.  

To generate the sequence of artificial observations $\mathcal{X}_0$ between $t=t_0$ and $t_w$ we then add, at each $t_i$, a vector of random numbers $\tilde{\bf e}$ (noise) to the true state:
\begin{linenomath*}\begin{equation}
	{\bf x}_{i,0}=\hat{\bf x}_i+\tilde{\bf e}_i
\end{equation}\end{linenomath*}
The random numbers are independently and identically distributed (IID) and are originally drawn from the Gaussian distribution $N(0,1)$. These random numbers are then converted to observational error by multiplying by a fixed fraction $\sigma$ of the natural variability $\bf r$ at that grid point,
\begin{linenomath*}\begin{equation}
	\tilde{\bf e}_i\sim\sigma\,{\bf r}\,N(0,1)
	\label{eq:obs-error}
\end{equation}\end{linenomath*}
The effect of the scaling by $\bf r$ ensures that, statistically, the amount of noise added at each point and to each field is the same fraction ($\sigma$) of the natural variability.

\subsection{Gradient descent applied to MORALS}
\label{sec:gd-morals}

To initialise the gradient descent algorithm, we first create a sequence of observations $\mathcal{X}_0$ of the true sequence $\hat{\mathcal{X}}$ using the method described in Sect.~\ref{sec:obs-seq} above.

The algorithm then enters its main loop, which advances the gradient descent by one iteration from step $h$ to $h+1$. The gradient descent loop begins with the sequence $\mathcal{X}_h$. The first step is to initialise MORALS using each state in the subsequence ${\bf x}_{i,h},\,i\in\{0,\ldots,w-1\}$, and then to integrate MORALS forward by $\Delta t=t_{i+1}-t_i$ to obtain the forecast images $f({\bf x}_{i,h}),\,i\in\{0,\ldots,w-1\}$. This is the \textit{forecasting step}. The mismatch is then calculated for each state-forecast pair, except for the first state where there is no corresponding forecast. The mismatch is given by
\begin{equation}
\delta{\bf x}_{i,h} = {\bf x}_{i+1,h}-f({\bf x}_{i,h})\equiv \left(\begin{array}{l}
	\delta{\bf u}_{i,h}\\
	\delta{\bf v}_{i,h}\\
	\delta{\bf w}_{i,h}\\
	\delta{\bf T}_{i,h}\end{array}\right)\equiv\left(\begin{array}{l}
	{\bf u}_{i+1,h}\\
	{\bf v}_{i+1,h}\\
	{\bf w}_{i+1,h}\\
	{\bf T}_{i+1,h}
	\end{array}\right)-f\left(\begin{array}{l}
	{\bf u}_{i,h}\\
	{\bf v}_{i,h}\\
	{\bf w}_{i,h}\\
	{\bf T}_{i,h}
	\end{array}\right)
	\label{eq:mismatch-morals}
\end{equation}
for $i\in\{0,1,\ldots,w-1\}$. The indeterminism of the sequence is then calculated, using a calculation based on Eq.~(\ref{eq:msi}). When applying this expression to MORALS, however, two problems must first be overcome. 

First, it is not appropriate to add together the different MORALS fields because they represent different quantities expressed in different units. To combine the quantities in this way they should be expressed in a non-dimensional form. This is true even for a non-physical quantity like the indeterminism because otherwise $I$ is poorly-defined - it depends on the units used for the different fields and its value can be changed just by changing those units even when the physical sequence of states itself has not changed. Second, the range of values in the four MORALS fields are quite different. $\bf T$ is usually $O(1)$, $\bf u$ and $\bf v$ are $O(10^{-2})$, and $\bf w$ is $O(10^{-3})$. The indeterminism combines the mismatch from all four fields, so the range of values in each field should be scaled before they are combined, otherwise the contribution to $I$ from the velocity mismatches will be swamped by the contribution from the temperature mismatches.

Both these problems are solved by dividing the mismatches grid point-wise by the natural variability $\bf r$ described in \ref{sec:obs-seq}. $\bf r$ remains constant over the course of the gradient descent, so the scaling is the same for each iteration. This solves the first problem by converting each value into a dimensionless quantity, and it solves the second by dividing by a natural scale for each field and at each grid point. With this scaling, we define the mean squared indeterminism for a sequence $\mathcal{X}_h$ to be
\begin{linenomath*}\begin{equation}
  I(\mathcal{X}_h) = \frac{1}{wN}\dsum_{i=0}^{w-1}\left\Vert\,\delta{\bf x}_{i,h}\circ{\bf r}^{-1}\right\Vert^2
  \label{eq:msi-morals-2}
\end{equation}\end{linenomath*}
where $\circ$ denotes the Hadamard (pointwise) product. The indeterminism may also be calculated for a particular ${\bf x}_{i,h}$:
\begin{linenomath*}\begin{equation}
	I({\bf x}_{i,h})=\frac{1}{N}\left\Vert\,\delta{\bf x}_{i,h}\circ{\bf r}^{-1}\right\Vert^2
	\label{eq:msi-onestate}
\end{equation}\end{linenomath*}
and for a single field, say temperature:
\begin{linenomath*}\begin{equation}
	I({\bf T}_{i,h})=\frac{1}{N_T}\left\Vert\,\delta{\bf T}_{i,h}\circ{\bf r}_{T}^{-1}\right\Vert^2
	\label{eq:msi-onefield}
\end{equation}\end{linenomath*}
where ${\bf r}_T$ denotes the temperature components of the $\bf r$ vector. Because each field is scaled in the same way, comparing these quantities between fields provides information about which fields are contributing most to the overall mismatch. As $I$ is a squared quantity, sums of these quantities also preserve the squared Euclidean norm:
\begin{eqnarray}
  I({\bf x}_{i,h}) & = & \frac{N_uI({\bf u}_{i,h})+N_vI({\bf v}_{i,h})+N_wI({\bf w}_{i,h})+N_TI({\bf T}_{i,h})}{N}\label{eq:indeterminism-state1}\\
  I(\mathcal{X}_h) & = & \frac{1}{w}\dsum_{i=0}^{w-1}I({\bf x}_{i,h})\label{eq:indeterminism-state2}
\end{eqnarray}
This definition of the indeterminism reflects its standard use by previous authors, as a mean over the model mismatches. The model grid itself is non-uniform; the model assigns higher grid resolution near the boundaries in order to resolve the boundary layers, but this is not reflected in the definition of the indeterminism as it should not be interpreted physically but as a purely mathematical construct. 

Once $I(\mathcal{X}_h)$ has been calculated it is compared with a user-specified value $\epsilon$. If $I(\mathcal{X}_h)\leq\epsilon$ then the gradient descent is terminated; $\epsilon$ is a parameter set by the user. If $I>\epsilon$ then a further check compares $I(\mathcal{X}_h)$ with $I(\mathcal{X}_{h-1})$. If $I(\mathcal{X}_h)>I(\mathcal{X}_{h-1})$ then the step length $\Delta\tau$ is halved, and the algorithm returns to the start of the \textit{previous} iteration: $h\to h-1$. If $I(\mathcal{X}_h)\leq I(\mathcal{X}_{h-1})$ then $\Delta\tau$ is doubled for the next iteration. This adaptive refinement of the step length allows the algorithm to proceed quickly both in regions of the state space where $I$ varies slowly with $\tau$ (adapting towards larger $\Delta\tau$) and where $I$ varies rapidly (adapting towards smaller $\Delta\tau$). In our runs we disable doubling of the step length once a point is reached when $I(\mathcal{X}_h)>I(\mathcal{X}_{h-1})$. The effect of this feature is that in the first few gradient descent steps $\Delta\tau$ equilibrates to a value that causes $I$ to fall smoothly in subsequent steps, while making optimum use of the available resources by running as few iterations as possible.

Finally $\mathcal{X}_{h}$ is updated. Using the gradient-free descent method of \citet{2004JuddB} with $\mathcal{A}=\lambda{\bf I}$, each new state is given by
\begin{equation}
	{\bf x}_{i,h+1}  =  {\bf x}_{i,h}-\frac{2\,\Delta\tau}{w}\times  \left\{\begin{array}{rcll}
	& - & \lambda\delta{\bf x}_{0,h} & i=0\\
	& & & \\
	\delta{\bf x}_{i-1,h} & - & \lambda\delta{\bf x}_{i,h} & 1\leq i\leq w-1\\
	& & & \\
	\delta{\bf x}_{w-1,h} & & & i=w
	\end{array}\right.
	\label{eq:gd-morals}
\end{equation}
where $\delta{\bf x}_{i,h}={\bf x}_{i+1,h}-f({\bf x}_{i,h})$ and $\lambda$ is a scalar. $\mathcal{X}_{h+1}$ is then the input for the next iteration of the gradient descent. The loop repeats while $I>\epsilon$ and $h\leq h_{\rm max}$, where $h_{\rm max}$ is a maximum iteration number.

\subsection{Significance level for the shadowing definition}
\label{sec:significance}

\begin{table*}
\centering
  \caption{Shadowing test parameters.}
  \begin{tabular}{lll}
	 \toprule
	 Best-case gradient descent & $\lambda$ & 0.25\\
	 Best-case iteration & $h$ & 9\\
	 Additional time & $t_{\rm end}-t_w$ & 1500\,s \\
	 Time of final state & $t_{\rm end}$ & 3920\,s \\
	 Number of Type I errors accepted & $R$ & 1\\
	 Number of candidate trajectories & $E$ & 130\\
	 Maximum number of significance tests in a trajectory & $n$ & 365\\
	 \bottomrule
  \end{tabular}
\end{table*}

We require a significance level $p$ for the shadowing definition such that Type I errors are avoided for all the model trajectories. The largest significance level $p$ such that in the event that the true candidate shadowing time is the trajectory length equivalent of $n$ significance tests, fewer than $R$ trajectories in a set of $E$ candidates will suffer a Type I error, is \citepalias[Eq.~14]{2019YoungQ}
\begin{linenomath*}\begin{equation}
  p=1-\left(1-\frac{R}{E}\right)^{1/2n}
  \label{eq:sidak}
\end{equation}\end{linenomath*}
We set $n$ to the longest possible model trajectory in this context, $n=365$. For the trajectory started from position $i=0$, $n=365$ comes from one significance test at lead time zero, $320/5=64$ tests over the sequence, and $1500/5=300$ over the extra period of observations. There are $E=129$ candidates (not including candidates started from forecast images, which use the same data), and to ensure fewer than one Type I error throughout the whole sequence we require $R<1$. Putting these into Eq.~\ref{eq:sidak} gives $p<10^{-5}$.

\subsection{Some comments on computational expense}

The computational resources required to run the gradient descent algorithm are considerable, even without a full adjoint model. The procedure was partially parallelized by running on a multi-core computer, with the forecast stage split into blocks of four simulations at a time. Even then each iteration of the gradient descent took approximately 90\,s (running on a single desktop computer with four Intel\textregistered\, Core\texttrademark\, 2 Q9400 CPUs running at 2.66GHz with 8GB RAM). 50--60\% of this time was spent on the forecast stage and 30--40\% setting up the parameter files for each simulation. Both of these steps require no cross-referencing from other simulations, so if run on a large cluster the computational overhead would be reduced significantly. The MORALS resolution is less than is normally used for simulations that are compared with laboratory data, which usually use $N_R=24$, $N_{\theta}=64$, $N_z=24$ or higher \citep{2008YoungA,2011Jacoby}. On a slightly older machine the $16\times32\times16$ run took 5\,min per iteration, a test run at $24\times64\times24$ required about 15\,min per iteration, and another test run at $8\times16\times8$ required only 90\,s. The medium resolution is sufficient for demonstration purposes and because we are working in the PMS, but if laboratory data were used to initialise the gradient descent then a higher resolution would be required.

\bibliographystyle{agufull08-ed}
\bibliography{012-013-LSE}

\begin{thebibliography}{40}
\providecommand{\natexlab}[1]{#1}
\expandafter\ifx\csname urlstyle\endcsname\relax
  \providecommand{\doi}[1]{doi:\discretionary{}{}{}#1}\else
  \providecommand{\doi}{doi:\discretionary{}{}{}\begingroup
  \urlstyle{rm}\Url}\fi

\bibitem[{\textit{Arakawa and Lamb}(1977)}]{1977Arakawa}
{\sc Arakawa, A., and V.~R. Lamb} (1977), {Computational Design of the Basic
  Dynamical Processes of the UCLA General Circulation Model}, \textit{Meth.
  Comput. Phys.}, 17:173--265, \doi{10.1016/B978-0-12-460817-7.50009-4}.

\bibitem[{\textit{Bowen}(1975)}]{1975Bowen}
{\sc Bowen, R.} (1975), {omega-Limit Sets for Axiom A Diffeomorphisms},
  \textit{J. Differ. Equations}, 18:333--339,
  \doi{10.1016/0022-0396(75)90065-0}.

\bibitem[{\textit{Du}(2009)}]{2009Du}
{\sc Du, H.} (2009), {Combining Statistical Methods with Dynamical Insight to
  Improve Nonlinear Estimation}, Ph.D. thesis, London School of Economics.

\bibitem[{\textit{Evensen}(1994)}]{1994Evensen}
{\sc Evensen, G.} (1994), {Sequential data assimilation with a nonlinear
  quasi-geostrophic model using Monte Carlo methods to forecast error
  statistics}, \textit{J. Geophys. Res.}, 99:10,143--10,162,
  \doi{10.1029/94JC00572}.

\bibitem[{\textit{Farmer and Sidorowich}(1991)}]{1991Farmer}
{\sc Farmer, J.~D., and J.~J. Sidorowich} (1991), {Optimal shadowing and noise
  reduction}, \textit{Physica D}, 47:373--392,
  \doi{10.1016/0167-2789(91)90037-A}.

\bibitem[{\textit{Farnell and Plumb}(1976)}]{1976Farnell}
{\sc Farnell, L., and R.~Plumb} (1976), {Numerical integration of flow in a
  rotating annulus II: three dimensional model}, \textit{Tech. rep.},
  Occasional Note Met O 21 76/1, Geophysical Fluid Dynamics Laboratory,
  Meteorological Office, Bracknell, Berkshire.

\bibitem[{\textit{Gilmour}(1998)}]{1998Gilmour}
{\sc Gilmour, I.} (1998), `$\iota$-shadowing, probabilistic prediction and
  weather forecasting', Ph.D. thesis, University of Oxford.

\bibitem[{\textit{Grebogi et~al.}(1990)\textit{Grebogi, Hammel, Yorke, and
  Sauer}}]{1990Grebogi}
{\sc Grebogi, C., S.~M. Hammel, J.~A. Yorke, and T.~Sauer} (1990), {Shadowing
  of physical trajectories in chaotic dynamics: Containment and refinement},
  \textit{Phys. Rev. Lett.}, 65:1527--1530, \doi{10.1103/PhysRevLett.65.1527}.

\bibitem[{\textit{Hammel}(1990)}]{1990Hammel}
{\sc Hammel, S.~M.} (1990), {A noise reduction method for chaotic systems},
  \textit{Phys. Lett. A}, 148:421--428, \doi{10.1016/0375-9601(90)90493-8}.

\bibitem[{\textit{Hide}(1953)}]{1953Hide}
{\sc Hide, R.} (1953), {Some experiments on thermal convection in a rotating
  liquid}, \textit{Q. J. Roy. Meteor. Soc.}, 79:161,
  \doi{10.1002/qj.49707933916}.

\bibitem[{\textit{Hide and Mason}(1975)}]{1975Hide}
{\sc Hide, R., and P.~Mason} (1975), {Sloping convection in a rotating fluid},
  \textit{Adv. Phys.}, 24:47--100, \doi{10.1080/00018737500101371}.

\bibitem[{\textit{Hignett et~al.}(1985)\textit{Hignett, White, Carter, Jackson,
  and Small}}]{1985HignettB}
{\sc Hignett, P., A.~A. White, R.~D. Carter, W.~D.~N. Jackson, and R.~M. Small}
  (1985), {A comparison of laboratory measurements and numerical simulations of
  baroclinic wave flows in a rotating cylindrical annulus}, \textit{Q. J. Roy.
  Meteor. Soc.}, 111:131--154, \doi{10.1002/qj.49711146705}.

\bibitem[{\textit{Hussain}(2010)}]{2010Hussain}
{\sc Hussain, M.} (2010), {Tangent Linear and Adjoint Models for Fluid Flow in
  a Rotating Annulus}, Master's thesis, Johann Wolfgang Goethe University.

\bibitem[{\textit{Ikeda}(1979)}]{1979Ikeda}
{\sc Ikeda, K.} (1979), {Multiple-valued stationary state and its instability
  of the transmitted light by a ring cavity system}, \textit{Opt. Commun.},
  30:257--261, \doi{10.1016/0030-4018(79)90090-7}.

\bibitem[{\textit{Jacoby et~al.}(2011)\textit{Jacoby, Read, Williams, and
  Young}}]{2011Jacoby}
{\sc Jacoby, T. N.~L., P.~L. Read, P.~D. Williams, and R.~M.~B. Young} (2011),
  {Generation of inertia-gravity waves in the rotating thermal annulus by a
  localised boundary layer instability}, \textit{Geophys. Astro. Fluid},
  105:161--181, \doi{10.1080/03091929.2011.560151}.

\bibitem[{\textit{Judd}(2003)}]{2003Judd}
{\sc Judd, K.} (2003), {Nonlinear state estimation, indistinguishable states,
  and the extended Kalman filter}, \textit{Physica D}, 183:273--281,
  \doi{10.1016/S0167-2789(03)00180-5}.

\bibitem[{\textit{Judd}(2007)}]{2007JuddA}
{\sc Judd, K.} (2007), {Failure of maximum likelihood methods for chaotic
  dynamical systems}, \textit{Phys. Rev. E}, 75:036,210,
  \doi{10.1103/PhysRevE.75.036210}.

\bibitem[{\textit{Judd}(2008)}]{2008Judd}
{\sc Judd, K.} (2008), {Forecasting with imperfect models, dynamically
  constrained inverse problems, and gradient descent algorithms},
  \textit{Physica D}, 237:216--232, \doi{10.1016/j.physd.2007.08.017}.

\bibitem[{\textit{Judd and Smith}(2001)}]{2001Judd}
{\sc Judd, K., and L.~Smith} (2001), {Indistinguishable states: I. Perfect
  model scenario}, \textit{Physica D}, 151:125--141,
  \doi{10.1016/S0167-2789(01)00225-1}.

\bibitem[{\textit{Judd and Smith}(2004)}]{2004JuddA}
{\sc Judd, K., and L.~A. Smith} (2004), {Indistinguishable states II. The
  imperfect model scenario}, \textit{Physica D}, 196:224--242,
  \doi{10.1016/j.physd.2004.03.020}.

\bibitem[{\textit{Judd and Stemler}(2009)}]{2009Judd}
{\sc Judd, K., and T.~Stemler} (2009), {Failures of sequential Bayesian filters
  and the successes of shadowing filters in tracking of nonlinear deterministic
  and stochastic systems}, \textit{Phys. Rev. E}, 79:066,206,
  \doi{10.1103/PhysRevE.79.066206}.

\bibitem[{\textit{Judd and Stemler}(2010)}]{2010Judd}
{\sc Judd, K., and T.~Stemler} (2010), {Forecasting: it is not about
  statistics, it is about dynamics.}, \textit{Philos. T. Roy. Soc. A},
  368:263--71, \doi{10.1098/rsta.2009.0195}.

\bibitem[{\textit{Judd et~al.}(2004)\textit{Judd, Smith, and
  Weisheimer}}]{2004JuddB}
{\sc Judd, K., L.~Smith, and A.~Weisheimer} (2004), {Gradient free descent:
  Shadowing, and state estimation using limited derivative information},
  \textit{Physica D}, 190:153--166, \doi{10.1016/j.physd.2003.10.011}.

\bibitem[{\textit{Judd et~al.}(2008)\textit{Judd, Reynolds, Rosmond, and
  Smith}}]{2008JuddA}
{\sc Judd, K., C.~A. Reynolds, T.~E. Rosmond, and L.~A. Smith} (2008), {The
  Geometry of Model Error}, \textit{J. Atmos. Sci.}, 65:1749--1772,
  \doi{10.1175/2007JAS2327.1}.

\bibitem[{\textit{Kostelich and Yorke}(1988)}]{1988Kostelich}
{\sc Kostelich, E.~J., and J.~A. Yorke} (1988), {Noise reduction in dynamical
  systems}, \textit{Phys. Rev. A}, 38:1649--1652,
  \doi{10.1103/PhysRevA.38.1649}.

\bibitem[{\textit{Lorenc et~al.}(1991)\textit{Lorenc, Bell, and
  Macpherson}}]{1991Lorenc}
{\sc Lorenc, A.~C., R.~S. Bell, and B.~Macpherson} (1991), {The Meteorological
  Office analysis correction data assimilation scheme}, \textit{Q. J. Roy.
  Meteor. Soc.}, 117:59--89, \doi{10.1002/qj.49711749704}.

\bibitem[{\textit{Lorenz}(1963)}]{1963Lorenz}
{\sc Lorenz, E.~N.} (1963), {Deterministic nonperiodic flow}, \textit{J. Atmos.
  Sci.}, 20:130--141, \doi{10.1175/1520-0469(1963)020<0130:DNF>2.0.CO;2}.

\bibitem[{\textit{Ravela et~al.}(2010)\textit{Ravela, Marshall, Hill, Wong, and
  Stransky}}]{2010Ravela}
{\sc Ravela, S., J.~Marshall, C.~Hill, A.~Wong, and S.~Stransky} (2010), {A
  realtime observatory for laboratory simulation of planetary flows},
  \textit{Exp. Fluids}, 48:915--925, \doi{10.1007/s00348-009-0752-0}.

\bibitem[{\textit{Rawlins et~al.}(2007)\textit{Rawlins, Ballard, Bovis,
  Clayton, Li, Inverarity, Lorenc, and Payne}}]{2007Rawlins}
{\sc Rawlins, F., S.~P. Ballard, K.~J. Bovis, A.~M. Clayton, D.~Li, G.~W.
  Inverarity, A.~C. Lorenc, and T.~J. Payne} (2007), {The Met Office global
  four-dimensional variational data assimilation scheme}, \textit{Q. J. Roy.
  Meteor. Soc.}, 133:347--362, \doi{10.1002/qj.32}.

\bibitem[{\textit{Read et~al.}(1992)\textit{Read, Bell, Johnson, and
  Small}}]{1992ReadA}
{\sc Read, P.~L., M.~J. Bell, D.~W. Johnson, and R.~M. Small} (1992),
  {Quasi-periodic and chaotic flow regimes in a thermally driven, rotating
  fluid annulus}, \textit{J. Fluid Mech.}, 238:599--632,
  \doi{10.1017/S0022112092001836}.

\bibitem[{\textit{Read et~al.}(2000)\textit{Read, Thomas, and
  Risch}}]{2000Read}
{\sc Read, P.~L., N.~P.~J. Thomas, and S.~H. Risch} (2000), {An evaluation of
  Eulerian and semi-Lagrangian advection schemes in simulations of rotating,
  stratified flows in the laboratory. Part I: Axisymmetric flow.}, \textit{Mon.
  Weather Rev.}, 128:2835--2852,
  \doi{10.1175/1520-0493(2000)128<2835:AEOEAS>2.0.CO;2}.

\bibitem[{\textit{Ridout and Judd}(2002)}]{2002Ridout}
{\sc Ridout, D., and K.~Judd} (2002), {Convergence properties of gradient
  descent noise reduction}, \textit{Physica D}, 165:26--47,
  \doi{10.1016/S0167-2789(02)00376-7}.

\bibitem[{\textit{Smith}(2000)}]{2000SmithA}
{\sc Smith, L.~A.} (2000), {Disentangling uncertainty and error: On the
  predictability of nonlinear systems}, in \textit{Nonlinear Dyanmics and
  Statistics}, edited by A.~Mees, pp. 31--64, Birkh{\"{a}}user Boston.

\bibitem[{\textit{Smith et~al.}(2010)\textit{Smith, Cu{\'{e}}llar, Du, and
  Judd}}]{2010Smith}
{\sc Smith, L.~A., M.~C. Cu{\'{e}}llar, H.~Du, and K.~Judd} (2010), {Exploiting
  Dynamical Coherence: A geometric approach to parameter estimation in
  nonlinear models}, \textit{Phys. Lett. A}, 374:2618--2623,
  \doi{10.1016/j.physleta.2010.04.032}.

\bibitem[{\textit{Stemler and Judd}(2009)}]{2009Stemler}
{\sc Stemler, T., and K.~Judd} (2009), {A guide to using shadowing filters for
  forecasting and state estimation}, \textit{Physica D}, 238:1260--1273,
  \doi{10.1016/j.physd.2009.04.008}.

\bibitem[{\textit{van Leeuwen}(2010)}]{2010vanLeeuwen}
{\sc van Leeuwen, P.~J.} (2010), {Nonlinear Data Assimilation in geosciences:
  an extremely efficient particle filter}, \textit{Q. J. Roy. Meteor. Soc.},
  136:1991--1999, \doi{10.1002/qj.699}.

\bibitem[{\textit{Young and Read}(2008)}]{2008YoungA}
{\sc Young, R. M.~B., and P.~L. Read} (2008), {Flow transitions resembling
  bifurcations of the logistic map in simulations of the baroclinic rotating
  annulus}, \textit{Physica D}, 237:2251--2262,
  \doi{10.1016/j.physd.2008.02.014}.

\bibitem[{\textit{Young and Read}(2013)}]{2013Young}
{\sc Young, R. M.~B., and P.~L. Read} (2013), {Data assimilation in the
  laboratory using a rotating annulus experiment}, \textit{Q. J. Roy. Meteor.
  Soc.}, 139:1488--1504, \doi{10.1002/qj.2061}.

\bibitem[{\textit{Young and Read}(2016)}]{2015Young}
{\sc Young, R. M.~B., and P.~L. Read} (2016), {Predictability of the thermally
  driven laboratory rotating annulus}, \textit{Q. J. Roy. Meteor. Soc.},
  142:911--927, \doi{10.1002/qj.2694}.

\bibitem[{\textit{Young et~al.}(2019)\textit{Young, Binter, and
  Nieh{\"{o}}rster}}]{2019YoungQ}
{\sc Young, R. M.~B., R.~Binter, and F.~Nieh{\"{o}}rster} (2019), {Shadowing
  the rotating annulus. Part I: Measuring candidate trajectory shadowing
  times}, \textit{arXiv}, physics.data-an:1909.04488.

\end{thebibliography}

\end{document}